\documentclass[authoryear, 5p, preprint]{elsarticle} 

\usepackage{subfigure}
\usepackage{threeparttable}
\usepackage{multirow}
\usepackage{lineno}
\usepackage{graphicx}
\usepackage{xcolor}
\usepackage{pdflscape}
\usepackage[utf8]{inputenc} 
\usepackage[T1]{fontenc}    

\usepackage{url}
\usepackage{amsmath}	
\usepackage{txfonts}
\usepackage{setspace}

\def \thetae {\theta_{\rm E}}

\def \zsource     {z_{\rm s}}

\def \zlens       {z_{\rm l}}

\def \sigmav      {\sigma_{\rm v}}

\def \beq {\begin{equation}}
\def \eeq {\end{equation}}
\def \beqn {\begin{equation*}}
\def \eeqn {\end{equation*}}

\newcommand{\Unif}{\ensuremath{\mathcal{U}}}

\def \figureSLbayesianres{
\begin{figure*}[!htp]

\centering
\includegraphics[width=0.9\textwidth]{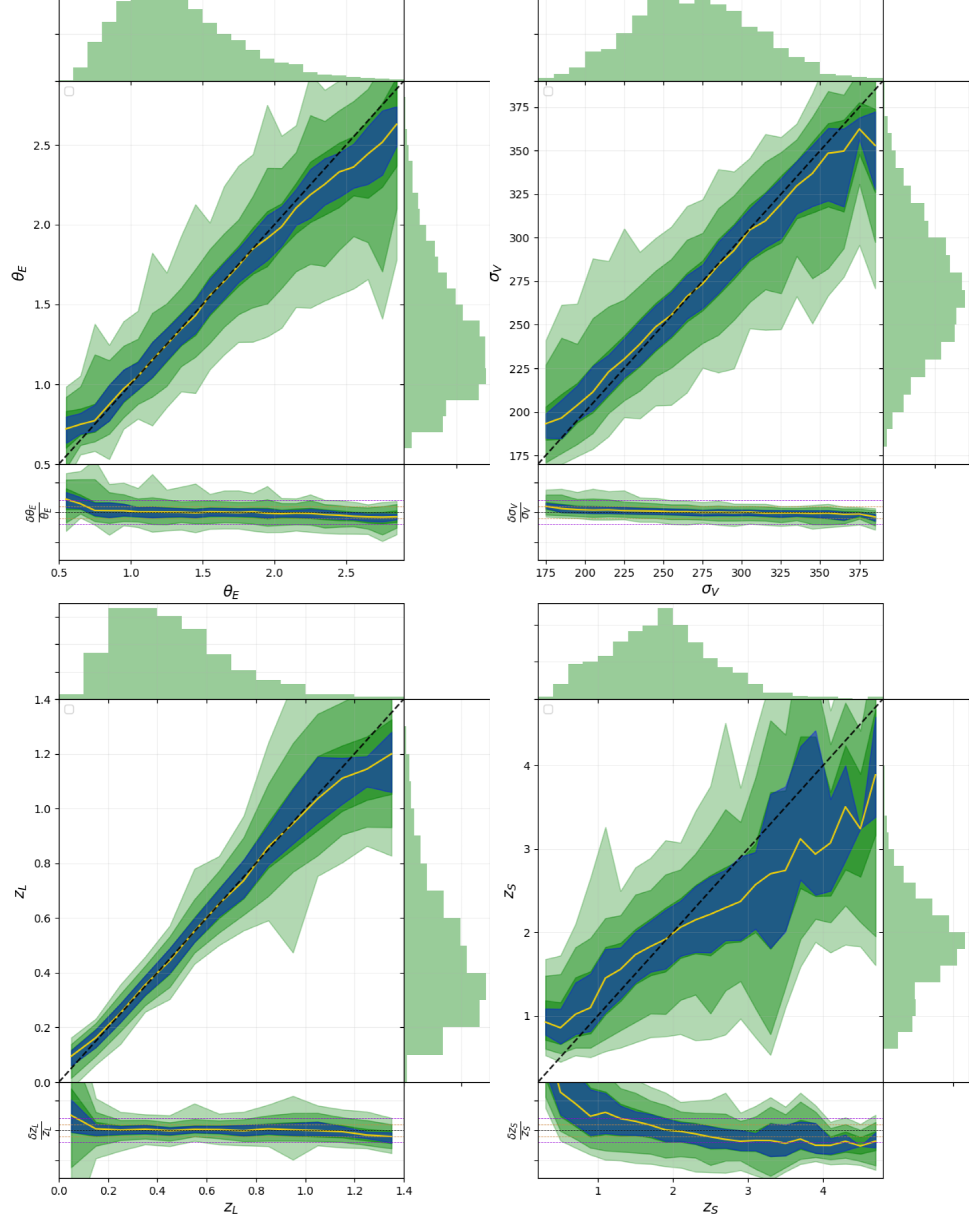}
    \caption{True values vs predicted values and the residuals for: the Einstein Radius,$\thetae$ (top-left), velocity dispersion, $\sigma_V$ (top-right), lens redshift, $z_L$ (bottom-left) and source redshift, $z_S$ (bottom-right). 
      The green shadows are the percentile corresponding to $1-2-3$ sigma area, the blue shadows are the 1-sigma scatter of medians predicted values.}
         \label{fig:SLres}

\end{figure*}

}

\def \figureSLvalidation{
\begin{figure*}[!htp]

\centering
\includegraphics[width=0.9\textwidth]{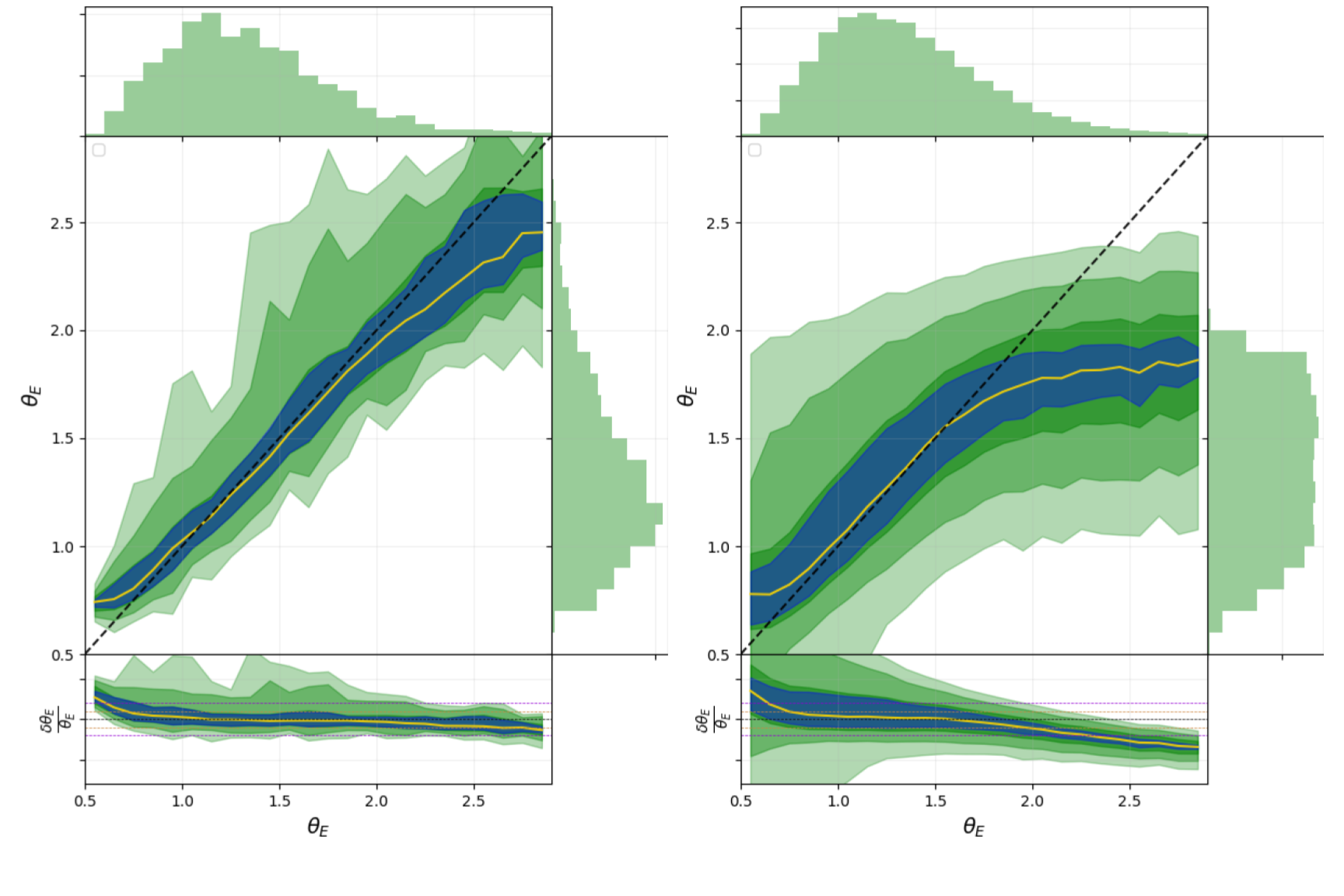}
    \caption{True vs predicted values of $\thetae$ and the residuals for a $90\%/10\%$ (left) $10\%/90\%$ training/test split.
      The green shadows are the percentile corresponding to $1-2-3$ sigma area, the blue shadows are the 1-sigma scatter of medians predicted values.}
         \label{fig:trainval}

\end{figure*}

}

\def \figuremodelinceptionA
{
\begin{figure}
   \centering
   \includegraphics[width=0.5\textwidth,keepaspectratio]{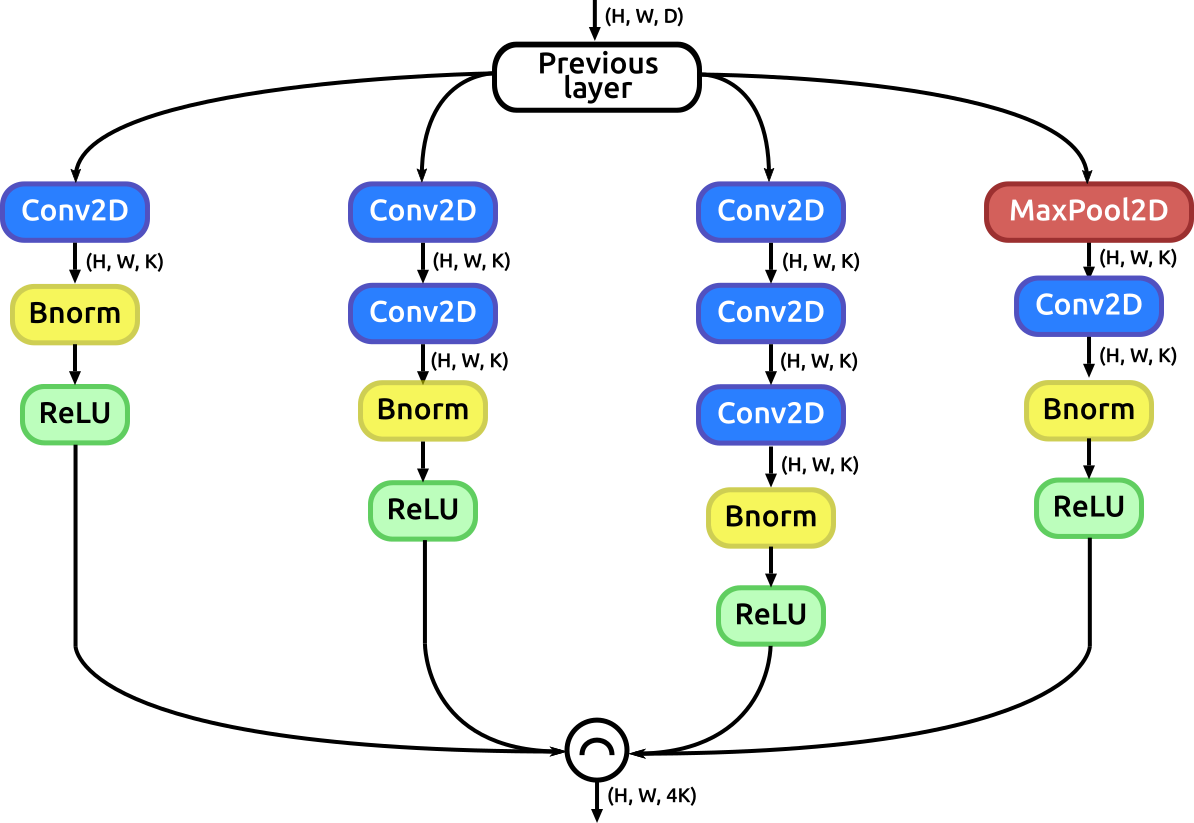}
      \caption{Diagram of the Inception block used in the Inception net, which is shown in Figure \ref{fig:Inception}. 
      This block is based on the inception module proposed by \cite{szegedy2016rethinking} for the Inception-v2 architecture. 
      This block allows the net to process data using different sizes of kernels, which is useful to enhance data features at different scales, without losing information by reducing its dimension. 
      The $\left(1,1\right)$ convolutions (the first convolutional layer appearing on each branch) are used here for image depth reduction (to make the training process less computational expensive). 
      The left branch only processes the input data using the $\left(1,1\right)$ convolutional scheme, to reduce the depth of the input data. 
      The second branch (from the left), has a  $\left(1,1\right)$ convolution followed by a $\left(3,3\right)$; while the third branch adds an extra $\left(3,3\right)$ convolution to the previous schema, thus making it possible to capture image features at greater scales. 
      Finally, the last branch uses a 2D max pooling layer to obtain translation invariance. 
      The activation maps of all branches are processed by a batch normalization and a ReLU activation at their respective ends. 
      Afterwards, the resulting maps are concatenated depthwise. 
      The dimension of the input data is $\left(H,W,D\right)$, and $K$ is a parameter previously specified (output channels).}
         \label{fig:InceptionA}
   \end{figure}
}

\def \figuremodeltraining
{
\begin{figure}
   \centering
   \includegraphics[width=0.4\textwidth,keepaspectratio]{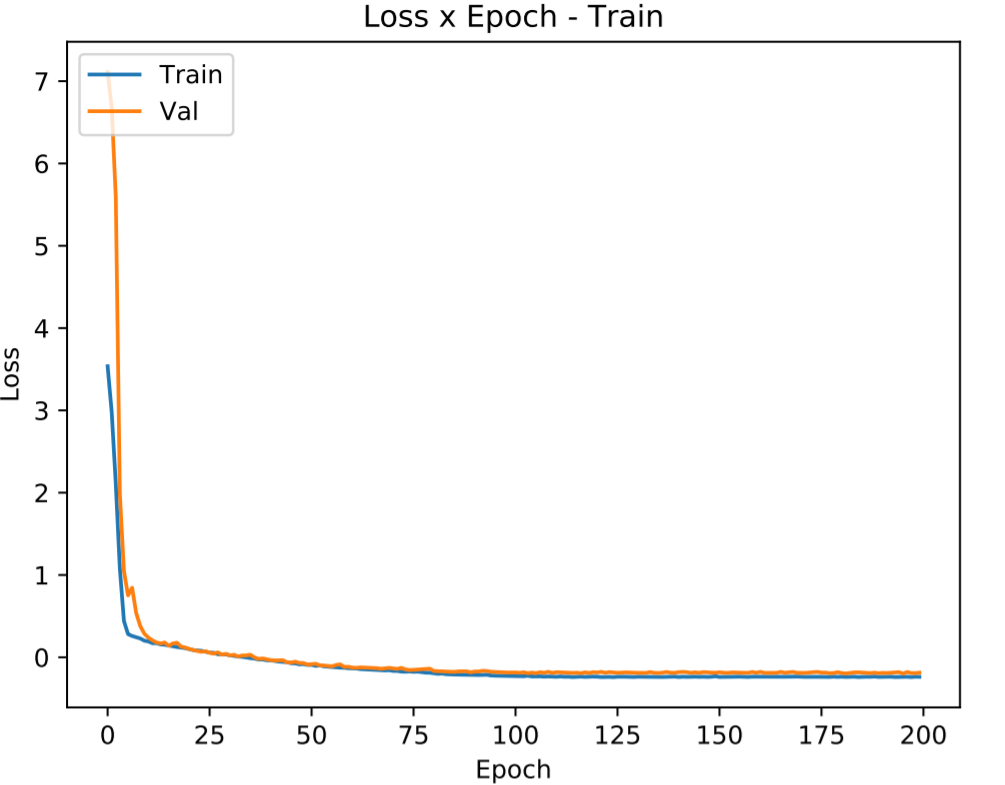}
      \caption{Loss versus epoch on training phase for $z_s$.}
         \label{training}
   \end{figure}
}

\def \figureoutputblock
{
\begin{figure}
    \centering
    \includegraphics[width=0.25\textwidth,keepaspectratio]{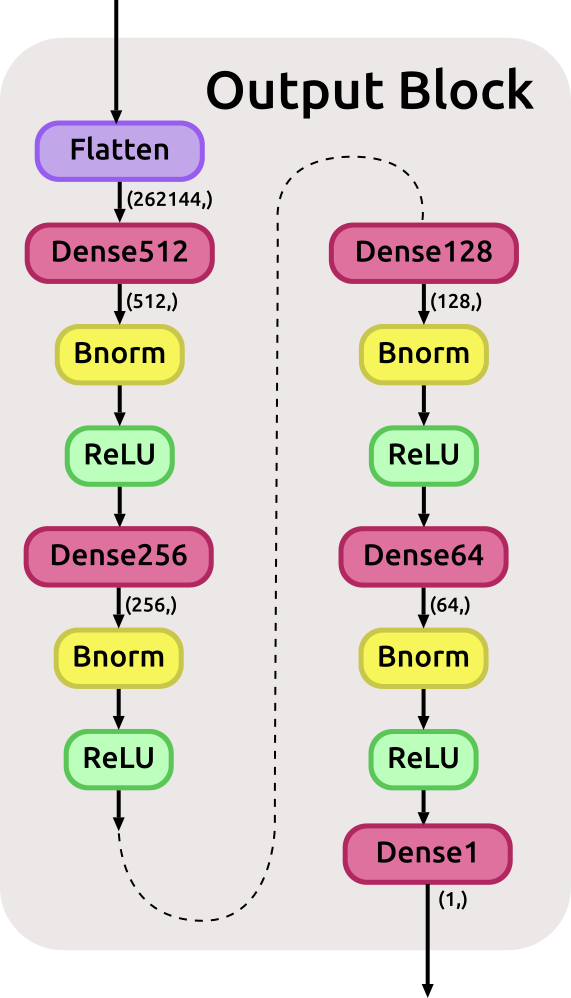}
    \caption{Diagram of the output block: a bottleneck-structured sequence — Dense/Batch Normalization /ReLU gradually decreasing feature hyperspace ($512$, $256$, $128$, $64$ and finally $1$).}
    \label{fig:output_block}
\end{figure}
}

\def \figuremodelinception
{
\begin{figure}
   \centering
   \includegraphics[width=0.5\textwidth,keepaspectratio]{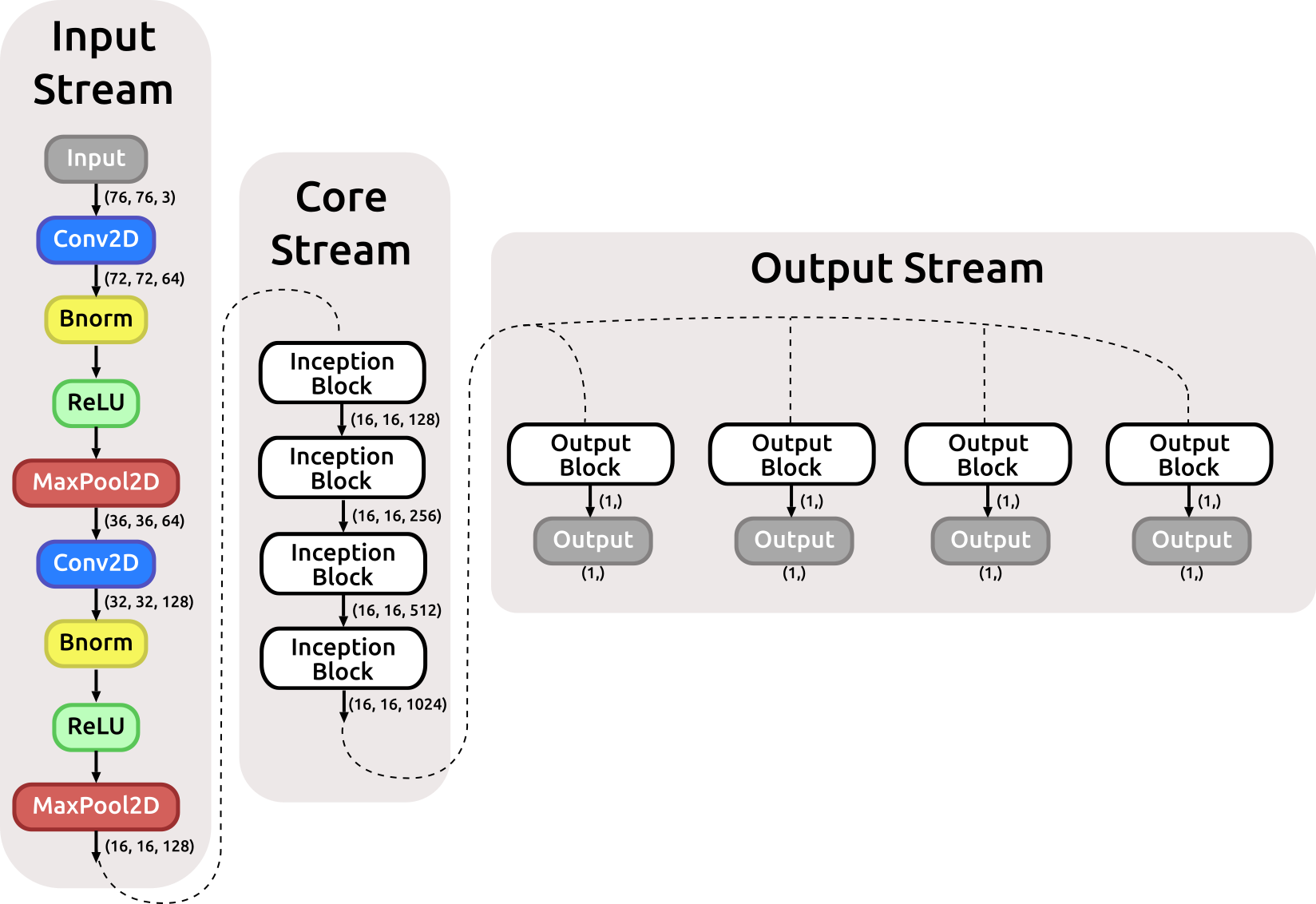}
      \caption{Diagram of the Inception network used in this work. 
      This architecture is composed of a total of 86 layers ( convolutional, batch normalization, ReLU activations, max pooling, and dense). 
      The size of the activation maps is shown at relevant locations of the net (after each layer that changes its size). 
      The architecture is divided into three main streams: the input stream, which is used to reduce dimensionality of input data and optimize training; the core stream, whose main purpose is to extract meaningful spatial features from the input data; and an output stream, which correlates these extracted features with each lensing property to be estimated (see Fig. \ref{fig:output_block}).}
         \label{fig:Inception}
   \end{figure}
}

\def \figurezlsigmain
{
\begin{figure*}
   \centering
   \includegraphics[width=0.9\textwidth,keepaspectratio]{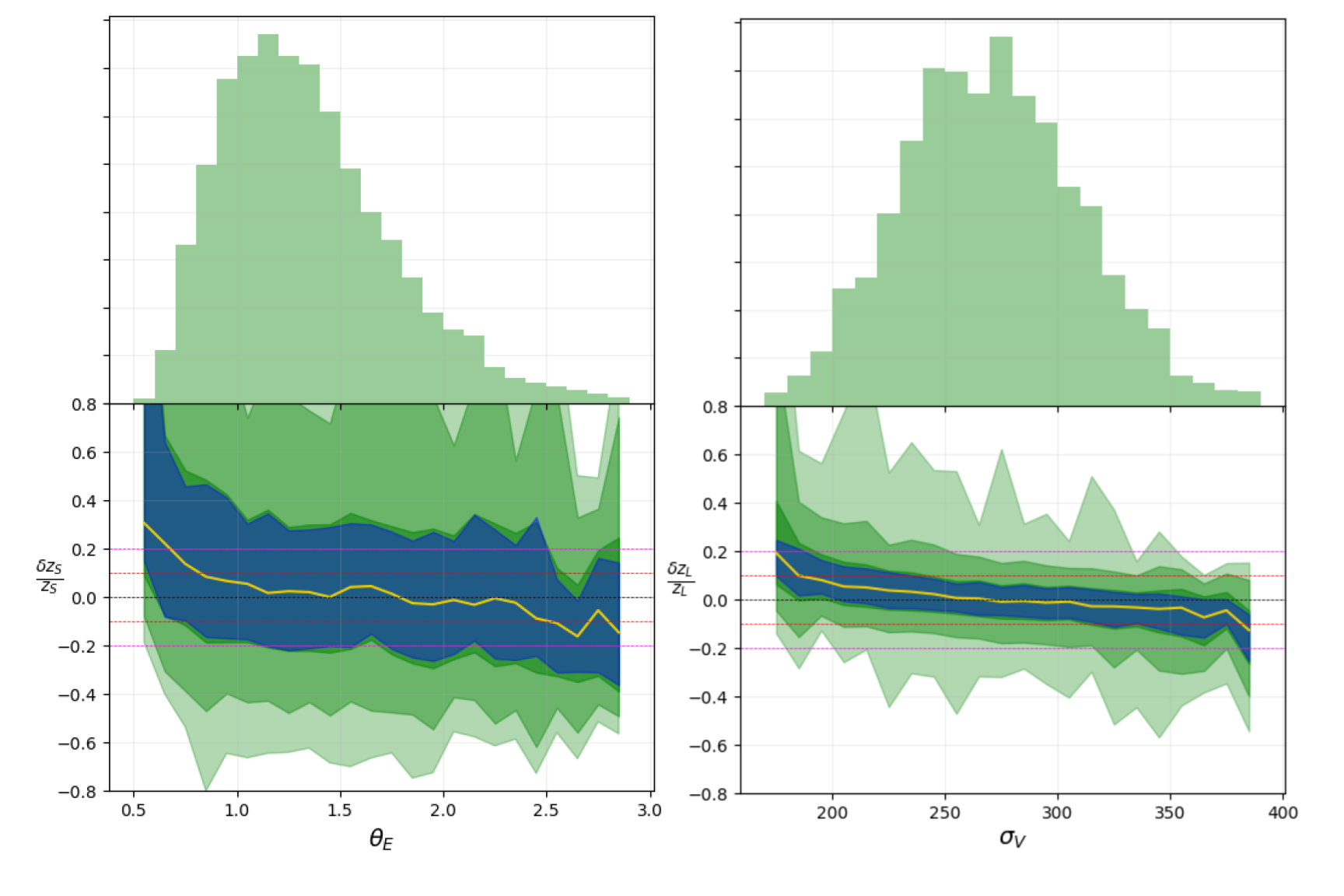}
      \caption{The fractional deviation from true values in
      source redshift $\zsource$ from as function of Einstein Radius $\theta_E$ (left) and  lens redshift $\zlens$ as function of velocity dispersion $\sigmav$ (right). 
      The green shadows are the percentile corresponding to $1-2-3$ sigma area and the blue shadows are the 1 sigma scatter of medians predicted values. The dashed lines represent the $0.0$ (black),$0.1$ (red),$0.2$ (magenta) deviations.}
         \label{fig:zl_sigma}
   \end{figure*}
}

\def \figureallin
{
\begin{figure}
   \centering
   \includegraphics[width=0.5\textwidth,keepaspectratio]{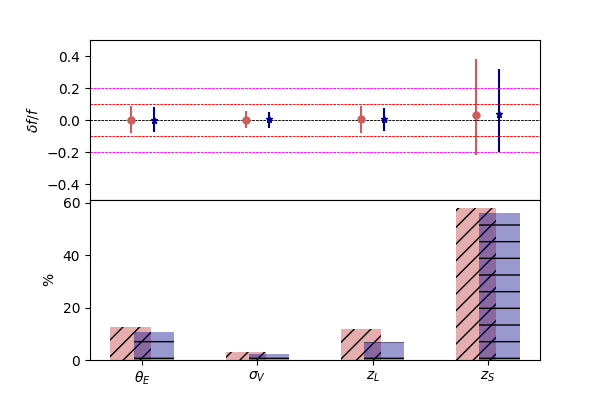}
      \caption{The fractional deviation on the full testing sample using models that predicts a single parameter (red circles) and the model that predicts all $4$ parameters at once (Blue star), where the dashed lines represent the $0.0$ (black),$0.1$ (red),$0.2$ (magenta) deviations (top). 
      The percentage of objects with more than $0.15$ fractional deviation (bottom). }
         \label{fig:sam_in}
   \end{figure}
}

\begin{document}

\begin{frontmatter}

\title{Deep Learning in Wide-field Surveys: Fast Analysis of Strong Lenses in Ground-based Cosmic Experiments}


\author[cfet,cbpf]{C.~R.~Bom\corref{cor1}} \ead{debom@cbpf.br}
\author[kicp]{J.~Poh}
\author[fnal,kicp,uofcaa]{B.~Nord}
\author[cbpf]{M.~Blanco-Valentin}
\author[cbpf]{L. O. Dias}

\address[cfet]{Centro Federal de Educa\c{c}\~ao Tecnol\'ogica Celso Suckow da Fonseca, Rodovia M\'ario Covas, lote J2, quadra J, CEP 23810-000,  Itagua\'i, RJ, Brazil}
\address[cbpf]{Centro Brasileiro de Pesquisas F\'isicas, Rua Dr. Xavier Sigaud 150, CEP 22290-180, Rio de Janeiro, RJ, Brazil}
\address[fnal]{Fermi National Accelerator Laboratory, P.O. Box 500, Batavia, IL 60510, USA}
\address[kicp]{Kavli Institute for Cosmological Physics, University of Chicago, Chicago, IL 60637, USA}
\address[uofcaa]{Department of Astronomy and Astropysics, University of Chicago, 5640 S. Ellis Ave., Chicago, IL 60134}

\begin{abstract}
Searches and analyses of strong gravitational lenses are challenging due to the rarity and image complexity of these astronomical objects. 
Next-generation surveys (both ground- and space-based) will provide more opportunities to derive science from these objects, but only if they can be analyzed on realistic time-scales. 
Currently, these analyses are expensive.
In this work, we present a regression analysis with uncertainty estimates using deep learning models to measure four parameters of strong gravitational lenses in simulated Dark Energy Survey data.
Using only $gri$-band images, we predict Einstein Radius ($\thetae$), lens velocity dispersion ($\sigmav$), lens redshift ($\zlens$) to within $10-15\%$ of truth values and source redshift ($\zsource$) to $30\%$ of truth values, along with predictive uncertainties. 
This work helps to take a step along the path of faster analyses of strong lenses with deep learning frameworks.
\end{abstract}


\end{frontmatter}

\section{Introduction}
\label{sec:introduction}

Strong gravitational lensing occurs when massive objects (e.g., galaxies and their dark matter haloes) deform spacetime, deflecting the light rays that originate at sources along the line of sight to the observer \citep[e.g.,][]{schneider2013gravitational,petters2012singularity,mollerach2002gravitational}. 
The key signature of a strong lens is a magnified and multiply imaged or distorted image of the background source, which can only occur if the source is sufficiently closely aligned to the line of sight of the warped gravitational potential generated by the lensing mass. 
Strong lensing also depends on the angular diameter distances between observer, lens, and source, which encloses information about the underlying cosmology. The source light may be magnified up to a hundred times, as the light deflection due to strong lensing conserves the surface brightness while increasing the total angular size of the object. 

Strong lensing systems are unique probes of many astrophysical and cosmological phenomena.
They act as ``gravitational telescopes,'' enabling the study of distant source objects that would be otherwise too faint to observe, such as high redshift galaxies \citep[e.g.,][]{2018ApJ...852L...7E,2011MNRAS.413..643R,2010MNRAS.404.1247J}:  dwarf galaxies \citep{Marshal2007}, star-forming galaxies \citep{Stark2008}, quasar accretion disks \citep{2008ApJ...673...34P}, and faint Lyman-alpha blobs \citep{2015arXiv151205655C}. 
Lensing systems can also be used as non-dynamical probes of the mass distribution of galaxies \citep[e.g.][]{0004-637X-575-1-87,2002MNRAS.337L...6T,2006ApJ...649..599K}, 
and galaxy clusters \citep[e.g.,][]{1989ApJ...337..621K,1998MNRAS.294..734A,2007MNRAS.376..180N,2010AdAst2010E...9Z,2010ApJ...715L.160C,2010ApJ...723.1678C}, providing a key observational window on dark matter \citep[see, e.g.,][]{2004MPLA...19.1083M}. 

Strong lensing has also been used --- alone or in combination with other probes --- to derive cosmological constraints on the cosmic expansion history, dark energy, and dark matter \citep[see, e.g.,][]{2010Sci...329..924J, GBC2016, 1998A&A...330....1B,1999A&A...341..653C,2002A&A...387..788G,2002MNRAS.337L...6T,2001PThPh.106..917Y,2004MPLA...19.1083M, 2005MNRAS.362.1301M,2010Sci...329..924J,Magana2015,0004-637X-806-2-185,GBC2016,Schwab2010,Enander2013,2016arXiv160203385P}.
Precise and accurate time-delay distance measurements of multiply-imaged lensed QSO systems have been used to measure the expansion rate of the universe \citep{2007ApJ...660....1O,2010ApJ...711..201S}. More recently, this technique has also been applied to multiply-imaged lensed supernovae \citep{2015Sci...347.1123K, 2016arXiv161100014G}. Strong lensing can also be used to constrain dark matter models \citep{2012Natur.481..341V, hezaveh2014measuring, gilman2018probing, rivero2018power, bayer2018observational}, as well as detect dark-matter substructures along the line-of-sight \citep{2018MNRAS.475.5424D, mccully2017quantifying}.

The broad range of applications has inspired many searches for strong lensing systems. 
Many of these searches have been carried out on high-quality space-based data from the Hubble Space Telescope (HST):  Hubble Deep Field \citep[HDF;][]{1996ApJ...467L..73H}, the  HST Medium Deep Survey \citep{1999AJ....117.2010R}, the  Great Observatories Origins Deep Survey \citep[GOODS;][]{2004ApJ...600L.155F}, the HST Archive Galaxy-scale Gravitational Lens Survey \citep[HAGGLeS;][]{marshall2009hst}, the Extended Groth Strip \citep[EGS;][]{Marshall2009robot}, and the HST Cosmic Evolution survey \citep[COSMOS;][]{2008ApJS..176...19F,2008MNRAS.389.1311J}.

However, there is also a plethora of ground-based imaging data that merits exploration. The majority of confirmed strong lensing systems that have been identified to-date were first discovered in ground-based surveys, such as the Red-Sequence Cluster Survey \citep[RCS;][]{2003ApJ...593...48G,2012ApJ...744..156B}, the  Sloan Digital Sky Survey \citep[SDSS;][]{2007Estrada,2009MNRAS.392..104B, 2010ApJ...724L.137K, 2011RAA....11.1185W,2012ApJ...744..156B}, the  Deep Lens Survey \citep[DLS;][]{Kubo2008}, The Canada-France-Hawaii Telescope (CFHT) Legacy Survey 
\citep[CFHTLS;][]{2007A&A...461..813C, 2012More,Maturi2014,RINGFINDER,SPACEWARPSII,ParaficzCFHTLS}, the Dark Energy Survey \citep[DES; e.g.,][]{DESSL1Nord}, the Kilo Degree Survey \citep[KIDS; e.g.,][]{2017arXiv170207675P}. Also, some Strong Lensing systems were firstly detected by Hershel Space observatory and South Pole Telescope (SPT) and later followed up by ALMA \citep{2010ApJ...719..763V, hezaveh2013alma,2012MNRAS.424.1614O,dye2018modelling,2010PASP..122..499E}. 
Strong lenses have also been discovered in follow-up observations of galaxy clusters \citep[e.g.,][]{1999A&AS..136..117L, 2003ApJ...584..691Z, 2008AJ....135..664H,2010A&A...513A...8K,2013MNRAS.432...73F} and galaxies \citep[e.g.,][]{2006MNRAS.369.1521W}. 
Next-generation surveys like LSST \citep{ivezic2008lsst}, Euclid \citep{laureijs2011euclid}, and WFIRST \citep{green2012wide}, are projected to discover up to two orders of magnitude more lenses than what is currently known \citep{2015ApJ...811...20C}. 

Many of the current catalogs of strongs lensing systems were found through visual searches.
However, the increasingly large data sets from current and future wide-field surveys necessitates the development and deployment of automated search methods to find and classify lens candidates.
Neural networks are one class of automated techniques. A number of recent works have demonstrated that both traditional neural networks \citep{2016arXiv160704644B, 2007Estrada} and deep neural networks \citep{2019MNRAS.484.3879P,2019MNRAS.484.5330J,2019MNRAS.482..807P, challenge,2018MNRAS.473.3895L, 10.1093/mnras/stx1492} can be used to identify morphological features in raw images that distinguish lenses from non-lenses, with minimal intervention from humans. 


In addition to finding catalogs of lenses, inferring the properties of lenses, like the Einstein Radius or the velocity dispersion of the lensing galaxy, typically require follow-up observations as well as modeling. 
Conventionally, modeling is performed with computationally expensive maximum likelihood algorithms \citep[e.g., ][]{2009ApJ...706.1201B,2005MNRAS.360..477D, 2008ApJ...681..814C, 2010ascl.soft10012O,2007NJPh....9..447J, 2014MNRAS.445.1942M,petkova2014glamer}, which can take up to weeks on CPUs and require manual input.
This is a relevant limitation for statistical studies of strong lenses or even for selecting systems to follow-up on. 
Recently, \citet{2017Natur.548..555H} showed that deep learning techniques could also be used in a regression task to produce fast measurements of strong lenses: the lens parameters in that work were measured on a set of high-quality HST simulations and images. Additionally, \citet{levasseur2017uncertainties} produced uncertainty estimates on strong lensing parameters using dropout techniques, which evaluates the deep neural network from a Bayesian perspective \citep{gal2016dropout}. The same approach was used by \citet{morningstar2018analyzing} to derive uncertainties in the parameters of strong gravitational lenses from interferometric observations.

In this work, we address the problem of strong lensing analysis in ground-based wide-field astronomical surveys, which have lower image quality than space-based data.
To develop and validate our deep neural network model, we produced a simulated catalog of strong lensing systems with DES-quality imaging.  
These simulations are used to train and evaluate the deep neural network model. The model is then used to infer the velocity dispersion $\sigma_v$, lens redshift $z_l$, source redshift $z_s$ and Einstein Radius $\theta_E$ of the lens. 
Our approach is generic: although the model is optimized for galaxy-scale lens systems (i.e., two objects, often visually blended, distorted image source, ring-like images or multiples images from the same source), it could be extended and optimized to analyze other species of strong lenses such as time-delay and double-source-plane systems.

This paper is organized as follows: 
First, in \S\ref{sec:deeplearningmodels}, we introduce the deep learning models and uncertainty estimation formalism used in this work.
Then, in \S\ref{sec:data}, we describe the simulated data used in this work.
Following that, in \S\ref{sec:training}, we describe how we trained the deep neural network model. In \S\ref{sec:results}, we apply our model to a test set and evaluate its performance.
Finally, we conclude and present an outlook for future work in \S\ref{sec:conclusion}.

\section{Regression with Uncertainty Measurements in deep learning Models}
\label{sec:deeplearningmodels}

Deep learning algorithms \citep{Goodfellow-et-al-2016,lecun2015deep}, and in particular Convolutional neural networks \citep[CNNs;][]{lecun1998gradient}, are established as the state of art for many sectors in computer vision research \citep[see, e.g.][]{ILSVRC15,zhang2018multi,lindeep2018,john2015pedestrian,peralta2018use}. In some cases, they have been shown to perform better than humans \citep[e.g.,][]{he2015delving,challenge}.
Deep learning allows the development of algorithms that can process complex and minimally processed (even raw) data from a wide variety of sources to extract relevant features which can be effectively linked to other properties of interest. 
For example, in computer vision tasks, it has been successfully applied to facial recognition \citep{lu2017simultaneous}, speech detection and characterization \citep{vecchiotti2018convolutional,abdel2014convolutional},  music classification \citep{choi2017convolutional}, medical prognostics \citep{li2018remaining} and diagnostics \citep{hannun2019cardiologist}. 

Deep learning models are not restricted to solve classification tasks, i.e. discrete-variable problems. 
Indeed, the the universal approximation theorem \citep{csaji2001approximation,Goodfellow-et-al-2016,hornik1991approximation,hanin2017universal, 2018arXiv180410306Y} states that a feed-forward neural network with a single hidden layer (depth-2)  with a suitable activation function can approximate a wide variety of continuous functions on compact subsets of $R^n$. In this case, the layer can be infeasibly large (wide). Recently, \citep{DBLP:journals/corr/abs-1709-02540} explored the width-bounded and depth-unbounded networks in which the authors argue that a width-$n+4$, where $n$ is the size of input layers with ReLU activation functions can approximate any Lebesgue integrable function on $n$-dimensional input space.
In both scenarios, depth-bounded or width-bounded,  these results  do not make statements on how this neural network can be trained.
Nevertheless, deep learning models have been successfully applied to estimate continuous variables, i.e. to perform a regression task \citep{lathuiliere2018comprehensive, lathuiliere2018deepgum, belagiannis2015robust}.

Regression problems can be formulated as a classification problem \citep{rothe2018deep,rogez2017lcr} if one discretizes the output parameters. 
For example, in deep learning for astronomy this approach has been recently applied to photometric redshift estimation \citep{pasquet2019photometric}. 
In that work, the algorithm predicts a set of parameters that represent the probability on each bin of the photometric redshift, which allows one to derive a probability density function (PDF).

However, this procedure faces several disadvantages. 
For example, the maximum precision of the probability peak region is limited by the bin size.
There is also a trade-off between the complexity and accuracy of the problem. 
In this scenario, during the optimization process, a high score in a bin next to the true value might have the same cost of a high score in a bin far from the true value. 
More sophisticated approaches can include multiple stages for deep regression, applying clustering, pseudo-labelling \citep{liu2016fashion}, or robust regression in which a probabilistic model, like Gaussian-uniform mixture, is added to make the model less sensitive to outliers \citep{lathuiliere2018deepgum}. 
Most of these models are usually very specialized and non-trivial to adapt for different sets of regression problems. 

In this work, we use a more direct and generic approach to adapt a typical classification deep learning model to the task of regression. 
We use an architecture based on the inception module \citep{szegedy2015going, szegedy2016rethinking}. Furthermore, in order to estimate uncertainties we apply the concrete dropout technique to approximate the PDF of the estimated values. 
Both the model architecture and the error estimations are detailed in next subsections.

\subsection{Model Architecture: Inception}
\label{sec:models}

Overly large neural network architectures pose a number of challenges in the training process. 
First, overfitting becomes a major limitation when the number of training samples do not scale as the number of parameters.
Additionally, a uniform increase in the number of filters in convolutional layers requires a quadratic increase in computation. 
Moreover, when new layers are added to large linear architectures, many weights become close to zero and most of the computation time is not fruitful \citep{szegedy2015going}.

The inception module (shown in Fig.~\ref{fig:InceptionA}) provides a non-linear architecture, as well as sparsity in the weights. 
The sparsity adds a relevant advantage by making the neural network more adaptable and stable. 
Additionally, a wider layer increases cardinality \citep{tishby2015deep,xie2017aggregated}, i.e., the number of independent paths which can provide a new way of adjusting the model.
With just thousands of parameters, the Inception module has been shown to outperform the traditional linear 
Visual Geometry Group Network \citep[VGG;][]{simonyan2014very} models that have tens of millions of parameters. 
Samples with objects of a variety of sizes present challenges for networks that lack the flexibility to contend with this.
Inception does not require a prescription for the optimal convolutional kernel size, because the convolutions are performed in parallel, each with a different kernel size.

In Fig.~\ref{fig:Inception}, we present the Inception architecture used in this work. 
Starting with the original architecture \citep{szegedy2015going}, we replace the regular multi-class softmax activation function \citep{krizhevsky2012imagenet} after the last dense layer with an unbounded linear activation that is able to output a single continuous value --- enabling the regression task.
It has three streams: the {\it input}, the {\it core}, and the {\it output}. 
The {\it input} stream reduces dimensionality. 
It is composed of two consecutive blocks of 2D convolutional layers, a batch normalization layer, a ReLU activation, and 2D max pooling layers. 
Both convolutional layers have a kernel size of $\left(5,5\right)$, while the pooling layers have a size of $\left(2,2\right)$.

\figuremodelinception

The core stream is composed of four consecutive Inception blocks. 
Each one of these blocks, whose structure can be seen in Fig.~\ref{fig:InceptionA}, has four branches, each with a different kernel size. 
The $\left(1,1\right)$ convolutions (the first convolutional layer appearing on each branch) are used for image depth reduction --- i.e., to lower computational cost. 
Any immediately following convolutional layer has a size of $\left(3,3\right)$\footnote{Note that stacking several $\left(3,3\right)$ convolutional layers is equivalent to single layers with greater size kernels. 
However, this stacking is more computationally efficient than using single kernels --- i.e., stacking two $\left(3,3\right)$ filters is equivalent to using a single $\left(5,5\right)$ kernel \citep{szegedy2016rethinking}}. 

As shown in Fig.~\ref{fig:Inception}, apart from its depth, the size of the internal activation maps inside the core stream is not modified. 
Other types of architectures need to reduce activation map dimensions to extract features at different scales. 
However, inception modules are expected to behave this way, keeping the dimensions of activation maps, as the features at different scales are extracted using the parallel convolutional schema as mentioned earlier.

\figuremodelinceptionA

After the Inception modules one needs to map the relevant features into the predicted variable. 
Therefore, we implemented a bottleneck-structured sequence --- Conv/BN/ReLU --- as proposed by \cite{he2016deep}.  We tested the current architecture in two different schemes: Building a model that predicts one parameter only, i.e., one trained model for each parameter independently,  and also a model that predicts all the four variables at the same time (see Fig. \ref{fig:output_block}). Besides the computational efficiency in the last approach we assure that the features that are used to predict $\theta_E$, for instance, are shared with the prediction of photometric redshift.

\figureoutputblock


\subsection{Error Analysis}
\label{sec:errors}

Uncertainty estimates are critical for assessing confidence in scientific measurements. 
Nevertheless, cogent and interpretable methods for uncertainty estimation in deep learning remain elusive. There are several types of uncertainty that are useful in assessing scientific confidence. 
They may be broadly classified into two categories --- {\it aleatoric} (statistical) and {\it epistemic} (systematic). 
Aleatoric uncertainties encompass effects that are unknown and change with the acquisition of each piece of data. 
These uncertainties are expected to decrease, for a given fixed model, with an increase in sample-taking in the predictions process.
For strong lenses, this would include shot noise in CCD imaging.
Epistemic uncertainties, on the other hand, include errors due to things that can be known but are neglect in the current investigation, for instance, certain effects that are not modelled.
For a given model these do not decrease with an increase in sample-taking. However, in the case of Deep Learning Regression which is a data driven model, if one feeds the network with more data during the training process that would change the model and, in principle could lower our ignorance about which model generated the collected training data \citep{kendall2017uncertainties}. We named the total error, which includes the epistemic and aleatoric, predictive error.
Standard error propagation is currently untenable, because there are not measurements of errors in raw images without performing some modeling in the first place. 
Additionally, there is no way to propagate that uncertainty, if it existed, through a deep learning model to the inferred parameters: we would need to know the uncertainties on the model parameters, but this error is not well known. 
Ideally, we would be able to perform uncertainty estimates of all parameters in a fully Bayesian framework.
Bayesian neural networks may provide Gaussian process approximations of the variance in the output parameters \citep{lee2017deep}; however, parameters may not all have Gaussian errors.

Another method that has been used recently is {\it Concrete Dropout}, which was first described by \cite{gal2016dropout,gal2017concrete}\footnote{github.com/yaringal/ConcreteDropout}, and first used in strong lens modeling by \cite{levasseur2017uncertainties}.
In this method, to estimate uncertainties on the parameters of interest, we compute the PDF of the predictions using the concrete dropout technique. 
Concrete dropout approximates a posterior distribution $p\left(Y|X\right)$ of the predicted physical parameter $Y$, given an input image $x$ in a Bayesian framework. 
We interpret our model in the variational perspective \citep{jordan1999introduction,NIPS2011_4329}.
We consider that dropping out neural network weights,i .e., performing dropouts, as a sampling from the distribution $p(\omega|{\bf X}, {\bf Y})$ of the weights $\omega$, which are learned via a set of inputs ${\bf X} = \{x_1, ...,  x_N \}$ and the corresponding output parameters ${\bf Y} = \{ y_1, ..., y_N \}$ \citep{gal2016dropout}. 

\subsubsection{epistemic (model) uncertainty}
Considering that neural networks can theoretically provide universal approximations, using dropout in this way is analogous to sampling over the space of functions \citep{gal2016dropout, gal2016uncertainty}.
The error associated with this sampling is related to the ignorance of the model; this is known as \textit{epistemic uncertainty}. 
Basics of method: a sampling a trained deep learning model could be interpreted in a Bayesian framework by optimizing its dropout rates and sampling the posterior $p( y| x, {\bf X}, {\bf Y})$ with the forward passes.
Then, once the network is trained, the sampling of predicted values are simply forward passes upon which we apply dropouts, and thus calculate a posterior for the predictions. 
This technique is known as Monte Carlo dropout \citep{gal2016dropout}.

However, defining the dropout rate is not a trivial task. 
For example, a fixed dropout probability will penalize larger weights when compared to the smaller ones \citep{gal2017concrete}, since when larger weights are dropped, they are likely to have a bigger impact in the results. 
To minimize the epistemic error in this scenario, one should optimize to lower-magnitude weights. For instance, the $0$-epistemic uncertainty would correspond to a situation in which we have all weights set to $0$, since the predictions would always be 0, however, the model would not perform any prediction at all. 
Therefore, the aim of defining a dropout rate is not to get optimized precision, but to find a point where epistemic errors can be reasonably defined. 
Some authors proposed a grid-search for this task. 
However, this procedure may be prohibitive in big and complex architectures. 

In the variational scheme, one may define a procedure to optimize the dropout rate.
The problem can be stated as follows: for a network, we can compute the PDF of a predicted value $y$ with input $x$ as:
\begin{equation}
p( y| x, {\bf X}, {\bf Y}) = \int p( y| x, \omega) \, p(\omega|{\bf X}, {\bf Y}) \, \, d\omega \, .
\label{posterior}
\end{equation}   
The posterior $p(\omega|{\bf X}, {\bf Y})$ has explicit dependence on the training dataset, and its form is generally infeasible to derive.
Thus we define an analytical variational distribution, $q_{\theta}(\omega)$, with parameters $\theta$ such that 
\begin{equation}
p( y| x) \approx \int p( y| x, \omega) \, q_{\theta}(\omega) \, \, d\omega \, .
\label{posterior_approx}
\end{equation}

\noindent We use a classification task as an example. 
From there, we develop a strategy for regression. 
For simplicity, in a classification task, it can be shown that one can derive the parameters from $q_{\theta}(\omega)$ by maximizing the log-evidence lower bound \citep{fox2012tutorial}:
\begin{equation}
L = \int q_{\theta}(\omega) \, \log p({\bf Y}|{\bf X}, \omega) \,\, d\omega - \mathrm{KL}(q_{\theta}(\omega)||p(\omega)) \, .
\label{objective}
\end{equation}
The first term corresponds to a traditional loss term in classification tasks --- i.e., a log-likelihood of the outputs for the training set --- which can be replaced with a Gaussian loss in regression tasks. 
The integral can be performed by a Monte Carlo integration procedure. 
The second term is $\mathrm{KL}$ divergence,
which parametrizes the distance between the distributions  $p(\omega)$ and $q_{\theta}(\omega)$,  thus minimizing it during the training process. The $\mathrm{KL}$ divergence term can be approximated by a $L_2$ regularization \citep{gal2016dropout}. For a set of parameters $\theta=\{{\bf M}, p_l\}_{l=1}^L$ in which ${\bf M}_l$ are the mean weight matrices and $p$ are the $l^{\rm th}$ layer, a typical choice for $q_{\theta}(\omega)$ is to define: 
\begin{equation}
\begin{split}
q_\theta(\omega) &= \prod_l q_{{\bf M}_{l}}({\bf W}_l), \\
q_{{\bf M}_l}({\bf W}_l)&={{\bf M}_{l}} \cdot \mathrm{diag}[\text{Bernoulli}(1-p_l)^{K_l}],
\end{split}
\label{q_parametrization}
\end{equation}
\noindent where the set of random weight matrices are $\omega = \{{\bf W}_{l=1}^L\}$,  with $L$ layers and dimensions of each weight matrix are $K_l$ and $K_{l+1}$.
The $\text{Bernoulli}$ variables, $z=\text{Bernoulli}(1-p_l)$, drop some neural network weights with its given probability. 

Thus, a deep learning model could be interpreted in a Bayesian framework by optimizing its dropout rates and sampling the posterior $p( y| x, {\bf X}, {\bf Y})$ with the forward passes. 
However, in some cases, there may be some issues in performing this optimization. 
It can be shown that finding the minimum of the $\mathrm{KL}$ divergence term in Eqn.~\ref{objective} is equivalent to maximizing the entropy of a Bernoulli random variable with probability $1 - p$. 
This penalizes larger models trained on small amounts of data, because it pushes the dropout rate close to $p=0.5$ in comparison to smaller models \citep{gal2017concrete}. 
Therefore, smaller models would have lower optimized dropout rates in the low-data regime. 
Nevertheless, with epistemic uncertainty, the dropout rate is lowered for both large and small model as we feed the neural network with more data.

There remain caveats when evaluating the derivative of the objective function with respect to a dropout rate in discrete Bernoulli distributions.
Therefore, we follow the prescription from \cite{gal2017concrete} and replace the Bernoulli variables for Concrete distribution \citep{maddison2016concrete} --- i.e., a continuous distribution with the ability to approximate discrete random variables. 
We sampled from the concrete distribution that approximates the one-dimensional Bernoulli, equivalent to a binary random variable\footnote{Note that that by the time \cite{gal2017concrete} article was published ,the method was not implemented to convolutional layers. 
We used the updated version in the aforementioned repository, which does work for convolutional layers.}:

\begin{align}
\tilde{z} &= \text{sigmoid} \left (
\frac{1}{t} \cdot \left(
\log p
- \log (1 - p)
+ \log u
- \log (1 - u)
\right)
\right),
\end{align}
\noindent where $t$ is a temperature parameter and $u$ is the uniform distribution $u \sim \Unif (0, 1)$.

After training, we derive $10^3$ realizations for each system. 
We define the $68\%$, $95\%$, and $99\%$ confidence intervals. 
We compared the confidence interval of the scatter of the medians --- scatter from different objects with same truth value --- with confidence levels from the individual object parameter realizations. 
The confidence intervals from the Concrete Dropout realizations were little wider but followed the scatter confidence intervals closely.  

\subsubsection{Aleatoric (statistical) uncertainties}
In principle, if one compares the results considering only epistemic error disregarding aleatoric errors to the truth values it might get unrealistic results. 
To address the aleatoric errors to the total (predictive) error one must take into account what are the noise proprieties of the dataset. 
For a \textit{homeostatic} data set --- in which all the data has similar noise proprieties --- this is usually done by adding a random uncertainty that can be manually fine-tuned. As the data presents diferent levels of signal-to-noise ratio we estimated the aleatoric uncertainties in a \textit{heteroscedastic} framework: we train the neural networks to predict the  $\sigma_k$, the observation noise parameter for the $k$ output parameter. This is done by optimizing $\sigma_k$ in the regression loss term, $L_R$ which corresponds to the first term of equation \ref{objective} for regression tasks and it is given by: 
\begin{align}
L_R = \sum_{k} \frac{-1}{2 \sigma_k^2}  ||{y}_{n,k} - \hat{{ y}}_{n,k}({\bf x}_n,\omega)||^2 - \frac{1}{2} \log {\bf \sigma}_k^2, 
\label{aleatoric}
\end{align}
where $y_{n,k}$ and $\hat{{ y}}_{n,k}$ are the true values and the predict values, respectively, for the $n$ training sampling. Thus, there is no need of labelled aleatoric uncertainties.


\subsubsection{Systematic uncertainties}
However, this still may not encapsulate all systematic uncertainties, which would be revealed in noise-free input data. Additionally, besides the source of epistemic (model) errors from the deep model uncertainty itself that might remain other degeneracies that can bias or scatter the results. For instance,
in wide-field survey imaging, the pixel size and PSF are typically larger than in space-based observations. 
There may also be degeneracies that can be more complex than random scatter on the predicted value. 
The Strong lensing Systems may have multiple source images that can be distorted in several ways, and can also be blended with the lens galaxy. Additionally, the lensed image has a parameter space with of order ten independent variables. This might be a relevant origin of systematic errors when trying to extract information from images. For example, it can be significantly easier to infer the Einstein radius, $\theta_E$, of a strong lensing system in cases where the light from the lensed source is not blended with the light from the lens galaxy than in cases where it is, even if the noise level in the images are the same. 

In order to evaluate a possible bias or scatter in our results, we visually compared the median predictions in our training sample to the respective truth values.
We observed that, even when considering the \text{epistemic uncertainties}, there was a small bias in some of our model predictions that scaled linearly. To address this problem, we adopted the following procedure: we performed a linear fit between model predictions and the truth values, and then subtracted the bias in the predictions. 
We then used the same linear fit to remove bias in the test data set. Therefore, our model comprises of a deep learning prediction and a linear scale correction.
After the fitting procedure, we found that the percentile error in the scatter in the medians and the percentile error due to the sampling performed by dropouts were consistently symmetric around the median and in most of the range around the $y=x$ line, except for certain high and low  boundaries that corresponded to regions where the model has fewer samples.
We discuss this further in section \ref{sec:conclusion}.

\section{Simulated Data}
\label{sec:data}

To optimally train, validate and test a neural network for strong lensing analysis, we require a large image catalog of strong lenses. Given the paucity of known strong lenses in the current census ($\sim$1000 lenses to date), we used simulated lenses from {\texttt LensPop}\footnote{\url{https://github.com/tcollett/LensPop}} \citep{2015ApJ...811...20C} to define different sets of images for training, validation and testing purposes. Here, we present a brief overview of the procedure used in the {\texttt LensPop} algorithm. For a complete description of {\texttt LensPop}, we refer interested readers to \cite{Collet2015}.

\texttt{LensPop} first generates a synthetic population of galaxy-scale strong lensing systems in the sky. For the lens population, \texttt{LensPop} assumes Singular Isothermal Ellipsoid (SIE) profiles for all lenses, with masses drawn from the velocity dispersion function of SDSS galaxies \citep{choi2007veldisp}. Observations show that elliptical galaxies, which dominate the lensing probability of the universe \citep[see, e.g.,][and references within]{Oguri2010} , are well-approximated by  SIE mass profiles \citep{Auger2010,Koopmans2006}. The redshift of the lenses are drawn independently from the mass from the differential comoving volume function. The light profile of the lens is assumed to follow a de Vaucouleurs profile \citep{1948AnAp...11..247D} that is aligned and concentric with the mass. The lens colors are assumed to follow the rest-frame SED of a galaxy whose star formation occurred 10 Gyrs ago. For the source population in \texttt{LensPop}, the source light have elliptical exponential light profiles, with magnitude, color and redshift distributions drawn from the sky catalogs of \cite{Connolly2010}.

The observing conditions of the imaging survey are then simulated and applied to the synthetic lenses to produce a mock catalog of lens imaging data that mimics that survey. In this work, we simulated the observing and instrumental capabilities of the DES survey to produce lenses with DES-like image quality.
The mock images are created by first pixelating the model lens image to the pixel scale of the detector of the survey instrument. 
The pixelated images are then convolved with circular atmospheric PSFs. 
Poisson noise from the lens, source, uniform sky background and CCD read noise are then added to the mock images. 
The zero-points, exposure-times, number of exposures, pixel-scale, read noise, filter bands and survey area are taken from DES survey specifications. The seeing and sky brightness are stochastic variables drawn from DES data and are described in Table 1 of \cite{Collet2015}.  

Every simulated lens in our data set is deemed DES-observable. 
We follow the criteria set in \cite{Collet2015} to determine which lensing systems are detectable by DES. 
All detectable lenses must be multiply imaged. 
Therefore we have:
\begin{equation}
    \thetae^2 > x_s^2 + y_s^2,
\end{equation}
\noindent where $\thetae$ is the Einstein radius, and $x_s$ and $y_s$ are the unlensed source position relative to the lensing galaxy. 
In at least one of the $g, r, i$ bands, the image and counter-image must be resolved. 
Hence, we also require that:
\begin{equation}
    \thetae^2 > r_s^2 + (s/2)^2,
\end{equation}
\noindent where $r_s^2$ is the half-light radius of the source, and $s$ is the seeing. 
Additionally, the tangential shear of the magnified source images in the image plane must also be resolved, and the magnification has to be large enough that the source images are noticeably sheared. 
Following \cite{Collet2015}, we adopt:
\begin{equation}
    \mu_{\mathrm{total}} r_s > s \quad\mathrm{and}\quad \mu_{\mathrm{total}} > 3 ,
\end{equation}
\noindent where $\mu_{\mathrm{total}}$ is the total magnification of the source. 
Finally, the signal-to-noise ratio, $S/N_{\mathrm{total}}$ must be high enough that it is feasible to identify the lens and to determine if the above criteria is met. 
Also following \cite{Collet2015}, we set 
\begin{equation}
    S/N > 20.
\end{equation}

\begin{figure*}[!htp]
\centering
	\subfigure{\includegraphics[width=\columnwidth]{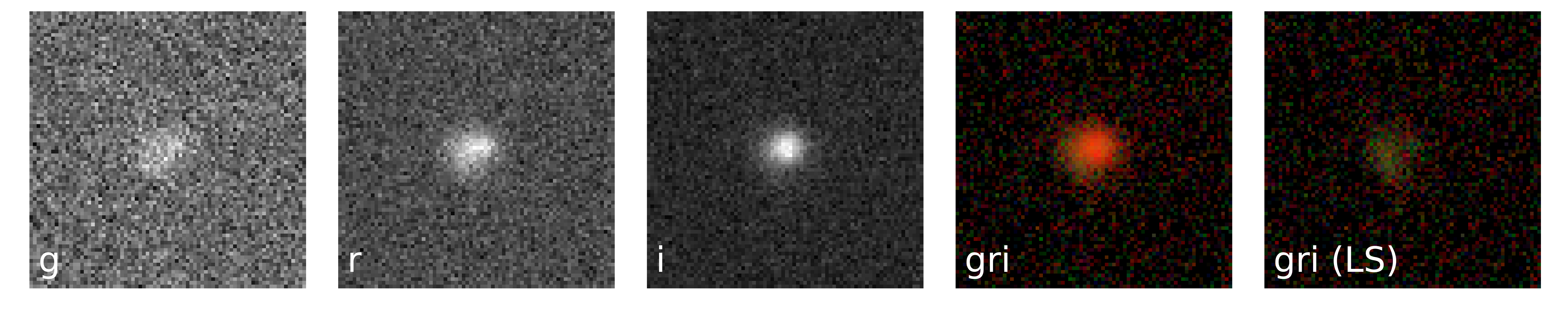}}
	\subfigure{\includegraphics[width=\columnwidth]{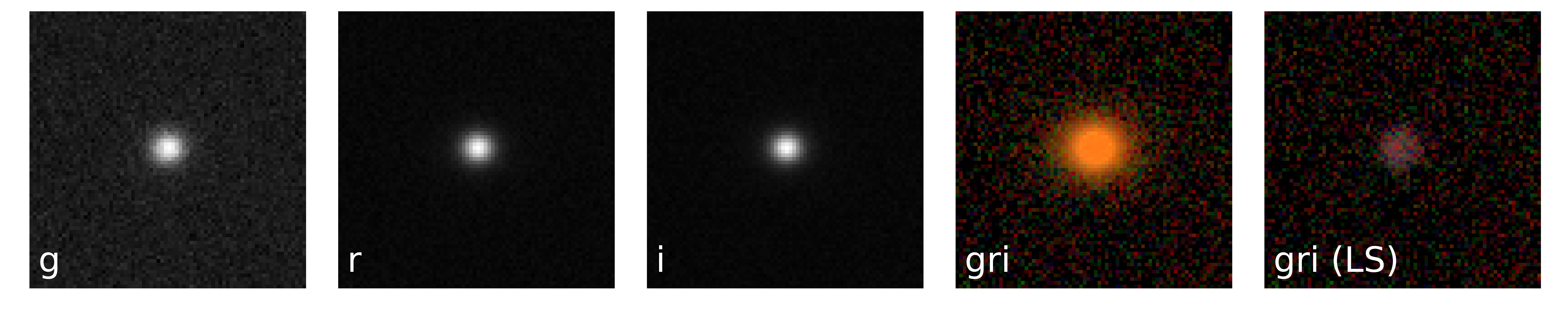}}
	\subfigure{\includegraphics[width=\columnwidth]{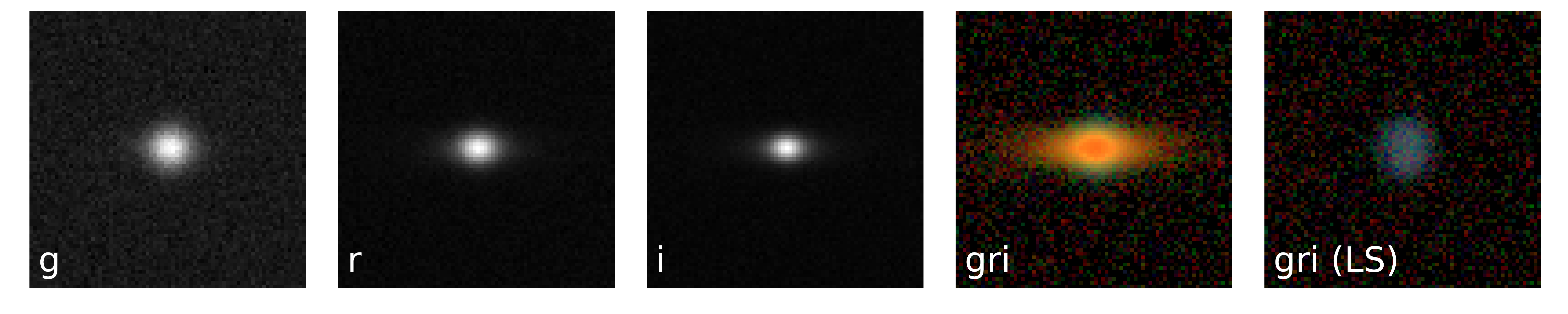}} 
	\subfigure{\includegraphics[width=\columnwidth]{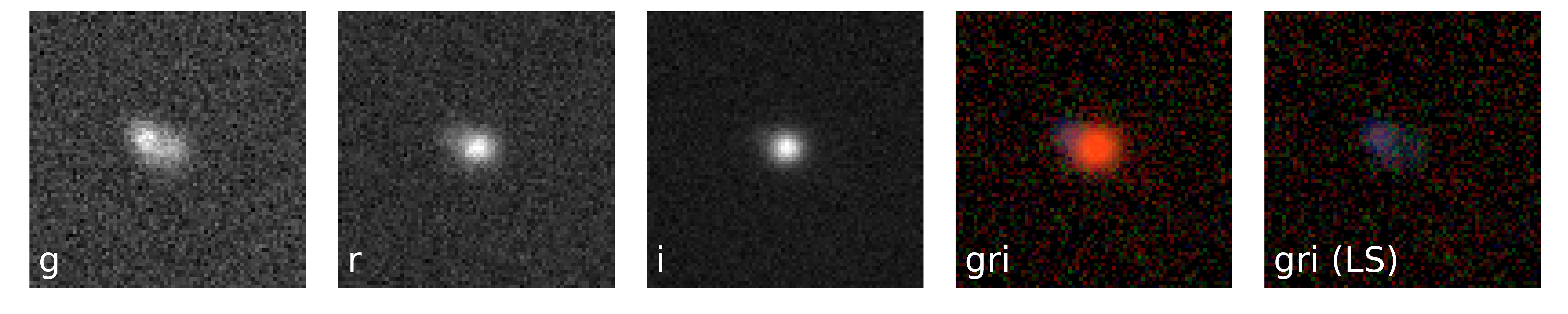}}
	\subfigure{\includegraphics[width=\columnwidth]{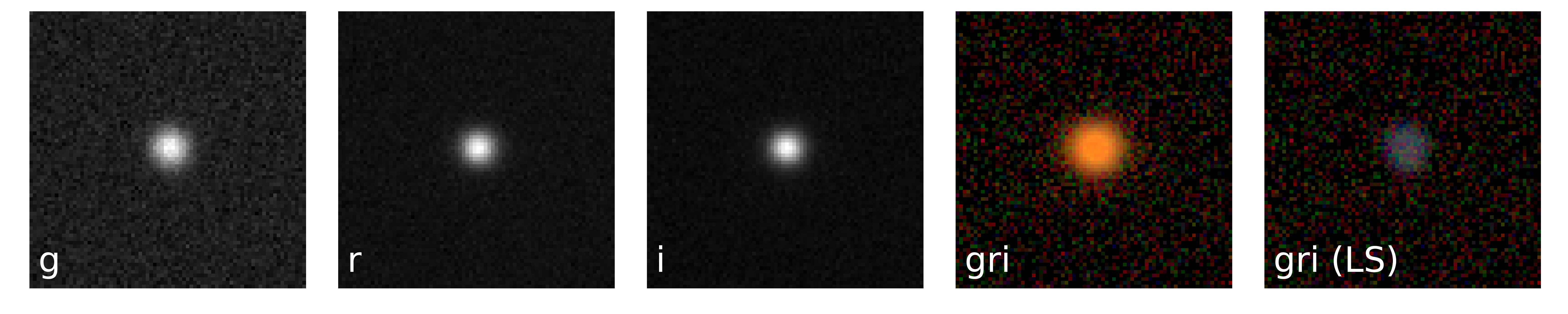}}
	\subfigure{\includegraphics[width=\columnwidth]{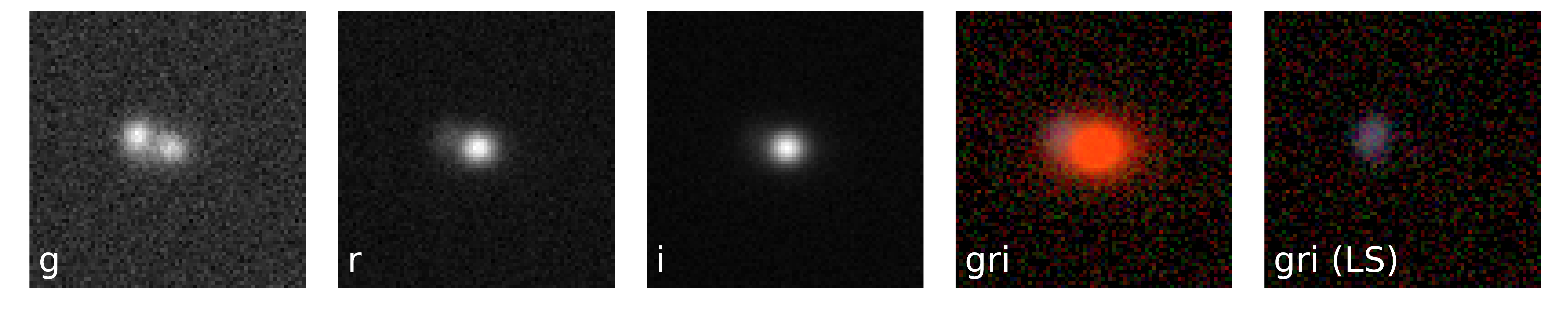}}
	\subfigure{\includegraphics[width=\columnwidth]{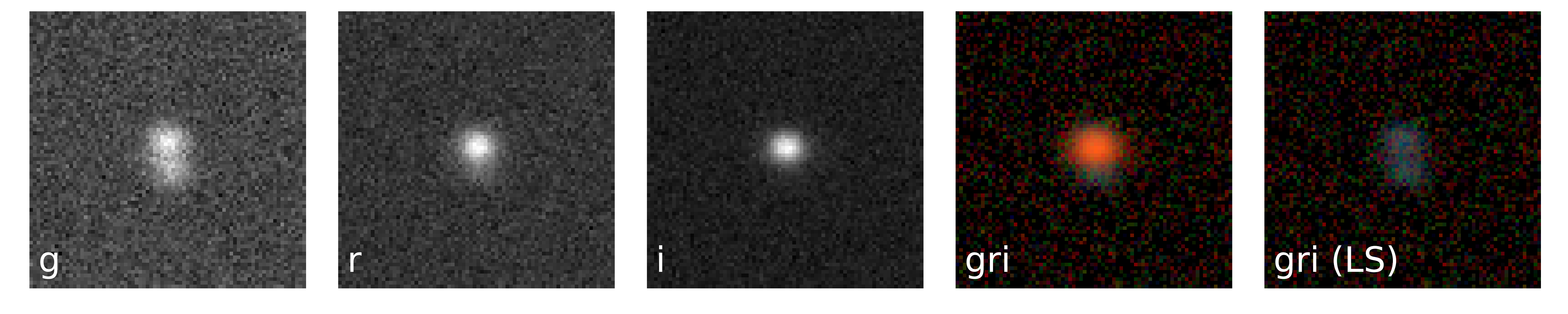}}
	\subfigure{\includegraphics[width=\columnwidth]{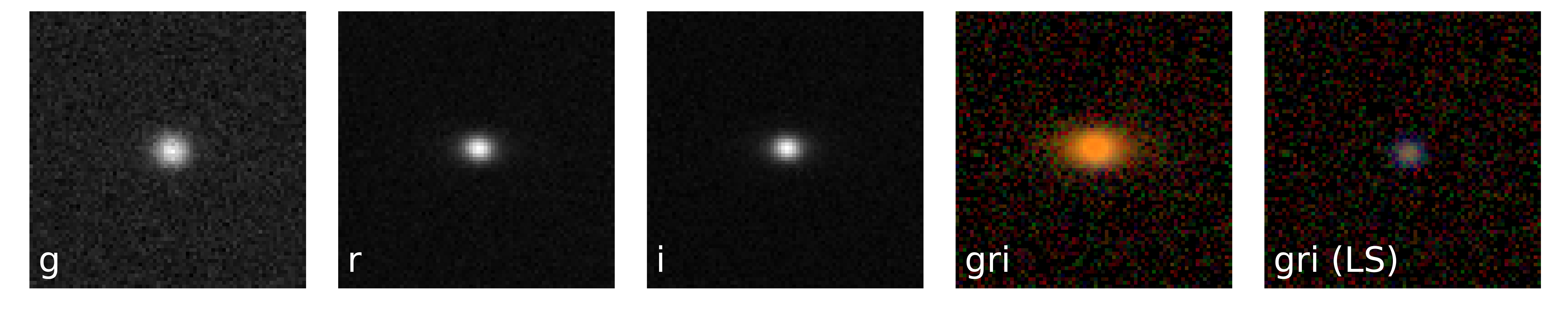}} 
	\subfigure{\includegraphics[width=\columnwidth]{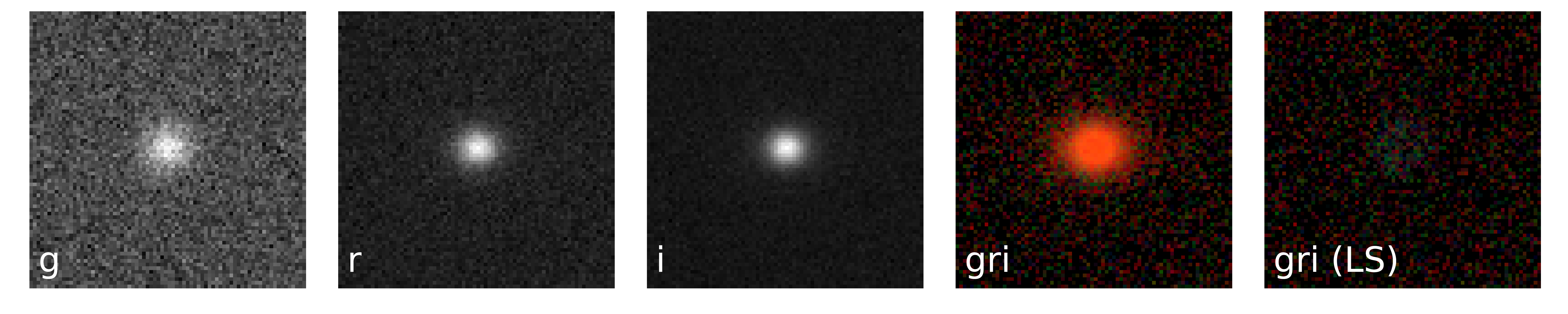}}
	\subfigure{\includegraphics[width=\columnwidth]{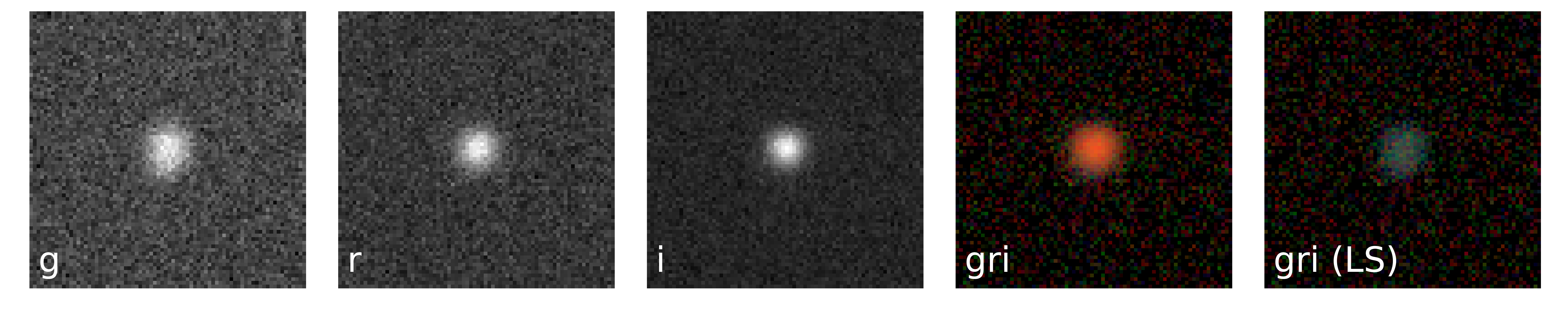}}
	\subfigure{\includegraphics[width=\columnwidth]{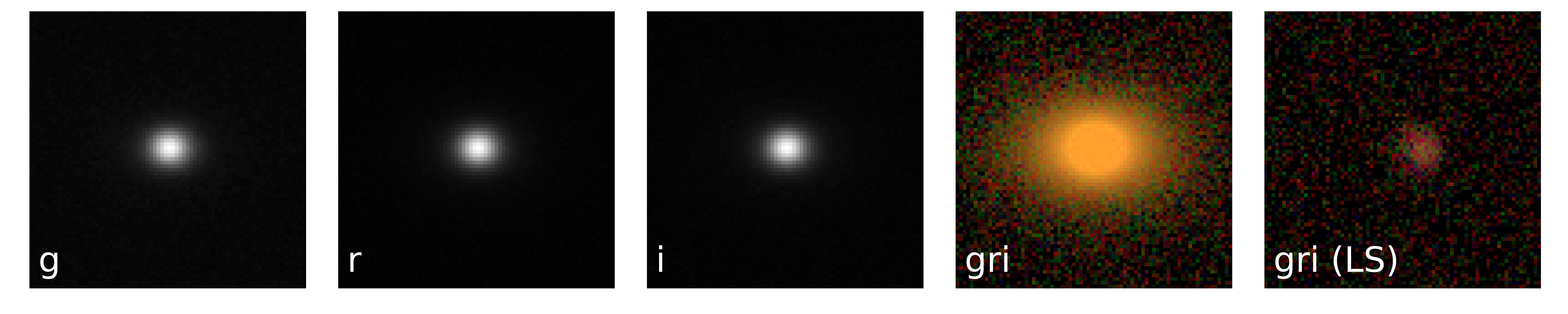}}
	\subfigure{\includegraphics[width=\columnwidth]{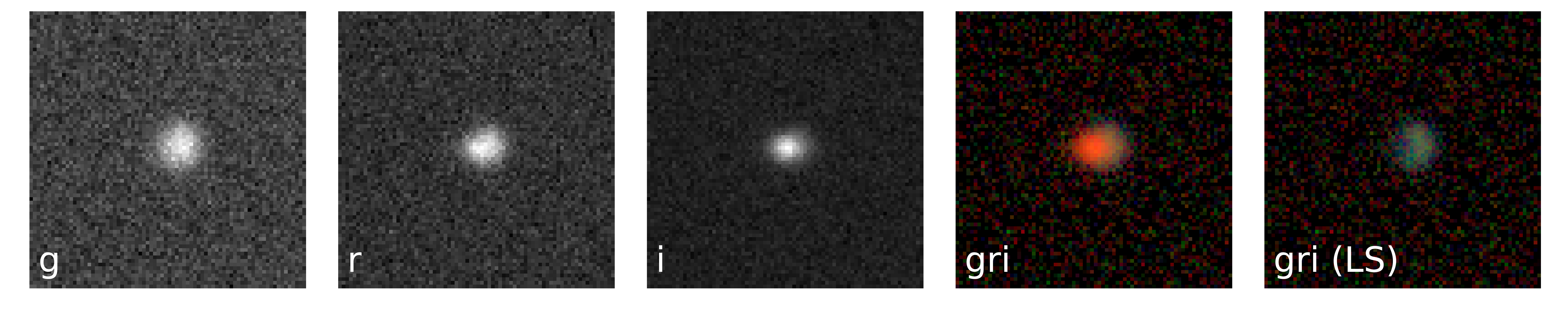}}
	\subfigure{\includegraphics[width=\columnwidth]{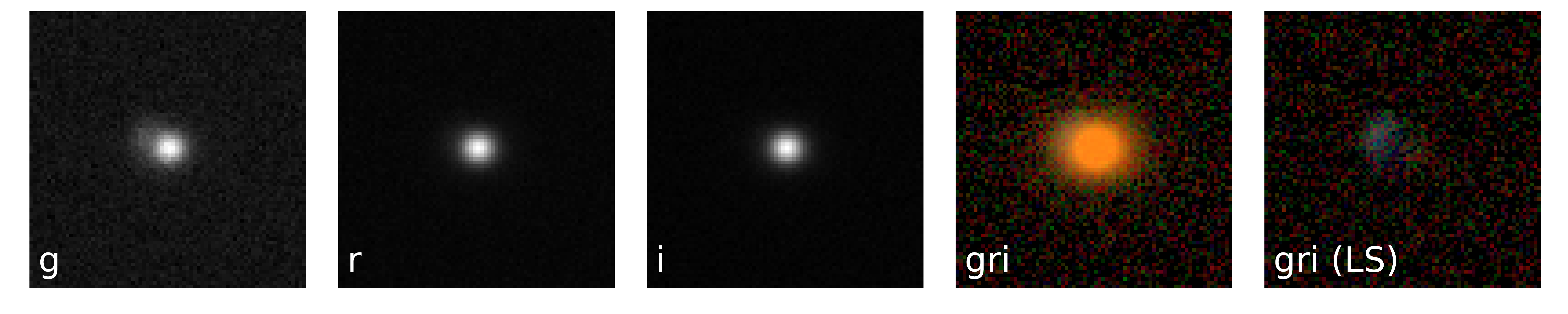}} 
	\subfigure{\includegraphics[width=\columnwidth]{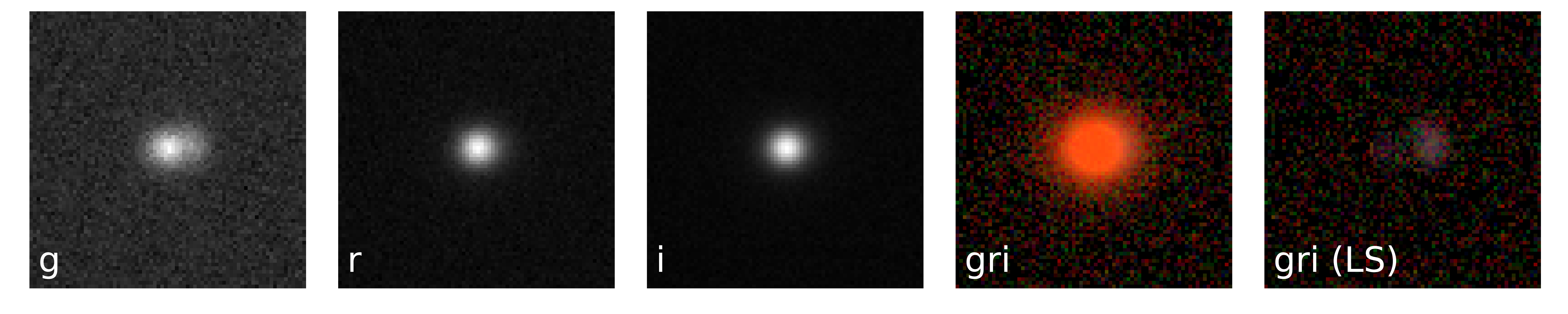}}
	\subfigure{\includegraphics[width=\columnwidth]{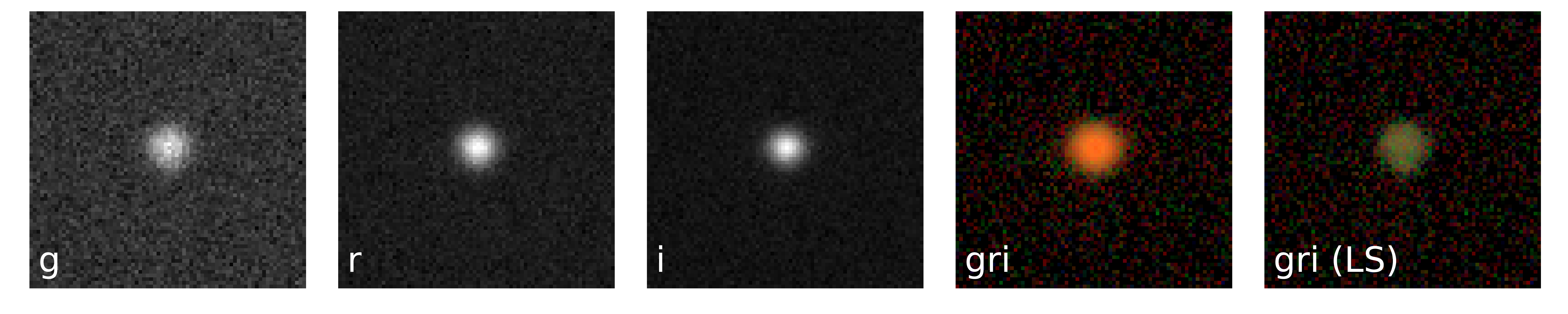}}
	\subfigure{\includegraphics[width=\columnwidth]{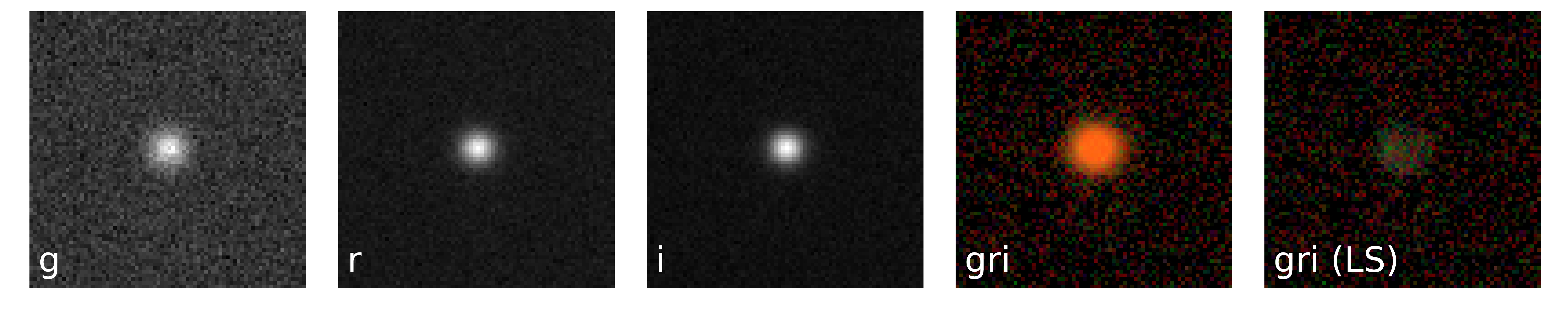}}
	\subfigure{\includegraphics[width=\columnwidth]{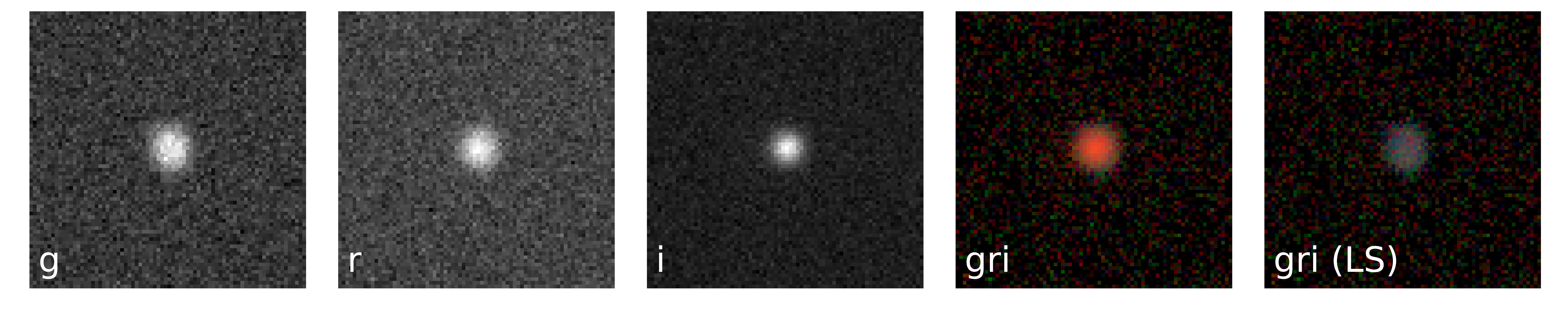}}
	\subfigure{\includegraphics[width=\columnwidth]{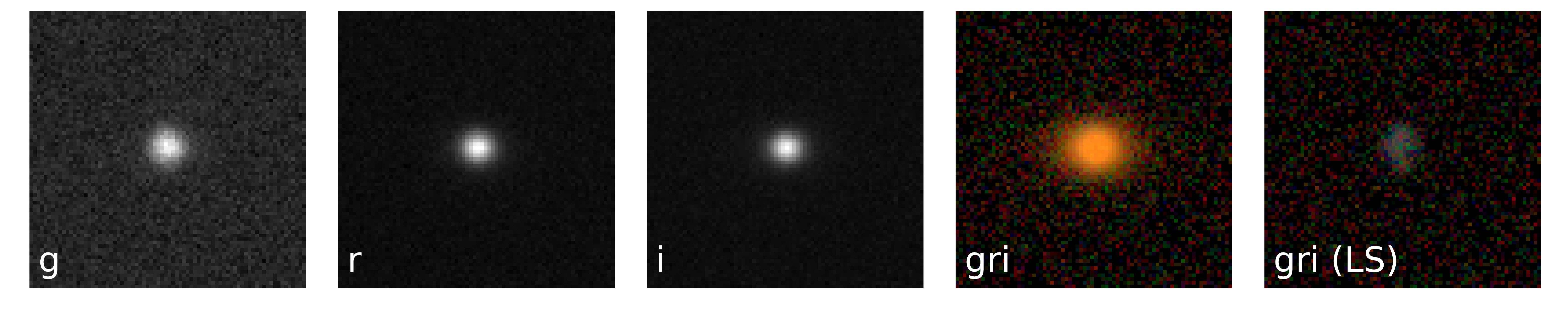}} 
	\subfigure{\includegraphics[width=\columnwidth]{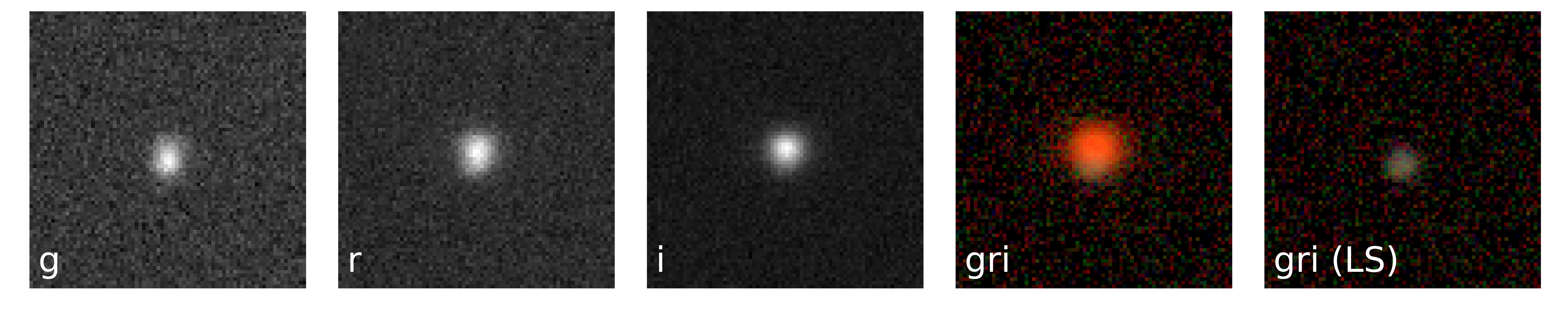}}
	\subfigure{\includegraphics[width=\columnwidth]{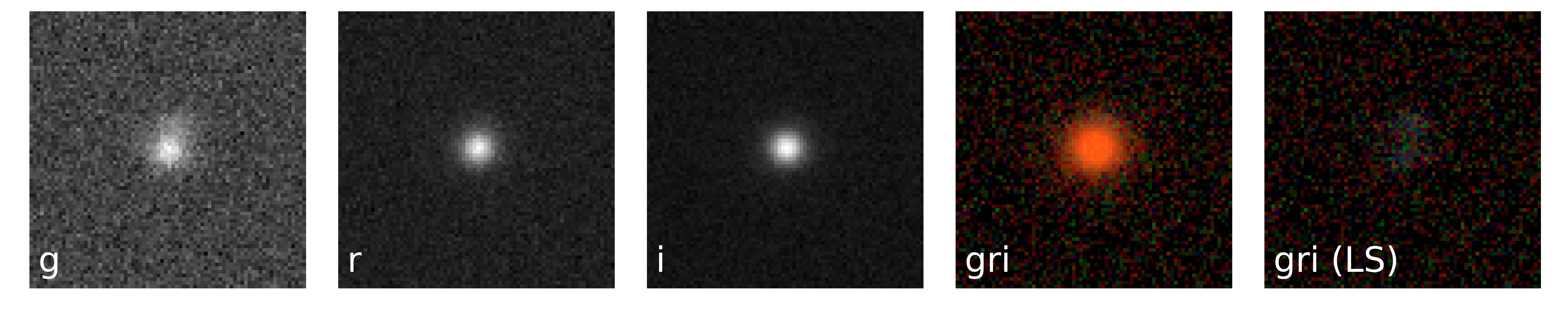}}
	\caption{\label{fig:sampleimages} A random selection of 20 DES-observable lens systems from the $\mathrm{LensPop}$ simulation. 
	For each system, the individual $g$, $r$, $i$, band images, as well as false-color $gri$ composite image with lens galaxy ('gri') and without lens galaxy ('gri (LS)'), are shown from left to right. 
	False-color images were made following \citet{2004PASP..116..133L}. 
	Images are sorted left-to-right, top-to-bottom by increasing Einstein radii. 
	Refer to \S \ref{sec:simulations} for further information on the simulated images dataset.}
\end{figure*}

Using \texttt{LensPop}, we generated $18,600$ simulated DES-observable galaxy-galaxy lensing systems. The distributions of the Einstein radii ($\thetae$), velocity dispersion ($\sigma_{\mathrm{v}}$), and lens and source redshifts ($z_L$, $z_S$) in the DES simulated dataset agrees with that of \cite{Collet2015}. Fig. \ref{fig:sampleimages} shows a representative random sample of 20 DES-observable systems from the total dataset. 

\section{Training the Inception Deep Learning Model}
\label{sec:training}

The strong lensing sample was divided into two groups: $80\%$ for training and validation purposes, and $20\%$ for testing. 
The training subset is the only one used to update the weights of the network in the backpropagation algorithm \citep{ruder2016overview}. 
We trained the Inception architecture for the parameters $\thetae$, $\sigmav$, $\zlens$ and $\zsource$  both together, all predictions at once and individually (in which the output core diagram presented in Fig.~\ref{fig:InceptionA} should be considered with only one branch).
The fine-tuned hyperparameters used for training the architecture were found by manually changing their values within a certain range until the (local) maximum accuracy on regression was achieved. 
The chosen batch size for training was $2,000$, while the maximum number of epochs was set to $400$. 
To avoid overfitting, and to improve model convergence, both a learning rate reducer and an early stopper were used.  
The training was performed with an Adam optimizer. 
The model was trained on a 24-core Intel~Xeon CPU X5670 ($2.93$ GHz) and a GeForce GTX 1080 GPU. 

The training time of each model was $\sim 4$ hours.
In Fig.~\ref{training}, we present the training diagnostic loss vs. epoch for the $z_{s}$, where we set a fixed number of epochs to $200$ for the training and validation dataset. 
The plot indicates the optimization performance on each sample, training and validation, and suggests no strong overfitting.
Both $\sigmav$, $\thetae$, $z_L$ and also the network that outputs all four parameter at once presented similar curves. 
\figuremodeltraining

\section{Results}
\label{sec:results}
The regression was performed with the trained models on our test sample consisting of $3720$ systems. 
The testing sample takes less than $\sim 1$ sec to infer $10^3$ realizations on each strong lens system. 
We present the predicted results for $\thetae$, $\sigmav$, $\zlens$ and $\zsource$ residuals relative to truth, along with the distributions of each value in Fig.~\ref{fig:SLres} in the network that predicts all four parameters at once.
We observed that, for most of the ranges, the predicted values remained within $\pm 10\%-15\%$ of the truth values, except for the $\zsource$. 
We noticed higher deviations from truth at low and high values, which correspond to smaller sample sizes in our training dataset. 
In those regions, some bias remains in the predicted medians and truth values, though they are consistent at $68\%$ confidence level.
In the top portion of Fig.~\ref{fig:sam_in}, we present the median of the fractional deviation and the $68\%$ confidence level percentile for both predicting one parameter at a time (red) and all parameters at once (blue).
We do not notice any strong bias in any parameter. 
In the bottom portion the same figure, we evaluate the size of the high-deviation sample, defined as fractional deviation higher than $15\%$). The results in both cases, red and blue, were similar, although it suggests that predicting several parameters at a time may lower the sample of high-deviation lenses at least for $z_L$.
For all parameters, except $\zsource$, the average fractional deviations in the whole testing sample was lower than $10\%$.

\figureSLbayesianres

\figureallin
\figureSLvalidation

We also investigated how the fractional deviation changes as function of other physical values, like magnitude, signal to noise ratio and size of the lensing object.
In most cases, we found regions of higher bias or error corresponding to low parameter sampling, suggesting that the uncertainty or bias could be overcomed with more data. 
However, there are some interesting cases, such as the ones presented in Fig.~\ref{fig:zl_sigma}. In these figures we observed that smaller/bigger $\thetae$ might be correlated to fractional deviations up to $\approx+80\%$/$-40\%$ considering the $68\%$ error bars for $\thetae > 0.7$ arcsec in the network that outputs all parameters at once. Though the error bars are wide and it includes the $0\%$ deviation, it is worth mentioning that there are regions in the high $\thetae$ end that have a lower number of examples and the errors might be poorly defined in those regions. The $\thetae$ is connected to $\zsource$ though the cosmological distances. In the $\zlens$ case we observe that lower/higher velocity dispersion may be linked to $\approx +20\%$/$-10\%$ deviations in the medians. Additionally to the presented plots, we also observed that $\thetae$ predicted errors are wider by a factor $\sim 2 $, i.e., $\sim  \pm 20\%$ for $\sigmav$ below $230$ km/s. As $\sigmav$ scales with the mass and therefore are connected to $\thetae$ these suggests that as the lensing effect gets weaker the uncertainty raises.These results might be useful if one is trying to define a more accurate sample or trying to fine tune the models. The results did not changed significantly in terms of accuracy as function of signal to noise ratio, this is probably due to the selection criteria $S/N >20$ in the simulations which requires that the strong lensing should be easy detected.
Lastly, we evaluated the effect of trainind/test dataset split. In the Fig.
\ref{fig:trainval} two different train/test sets with $90\%/10\%$(left) and $10\%/90\%$ (right) are shown. The right figure presents wider errorbars, e.g., for $\thetae<1.0''$ the high limit of $3$ sigma uncertainty is $2.0''$. It is worth noticing that the methods became flat, considering the medians for $\thetae \gtrapprox 2.0''$. These results support the importance of using as much data as possible in data driven models such as the one presented in order to make reasonable predictions.

\figurezlsigmain

\section{Discussion and concluding remarks}

\label{sec:conclusion}
We presented the prediction of astrophysical features of strong lensing sytems in simulated wide-survey images using a deep neural net model. 
These parameter predictions include estimates of both epistemic and aleatoric uncertainties, and we verified that the scatter on thousands of individual systems predictions were lower than this uncertainty. 
In particular, the velocity dispersion was constrained to lower than $10\%$ level using only $3$ bands.
The current results support that we could use deep learning as a tool for quick catalog generation and a reliable analysis that could outperform more conventional methods (e.g., MCMC) in computation time and without highly specialized experts. 
This, in principle could be used to select systems for further investigation or be used in statistical analysis that requires this $\sim 15\%$ level performance, for instance in galaxy-galaxy strong lensing cosmology \citep{0004-637X-806-2-185,chen2018assessing}, where we could use the $\thetae$ combined with an independent measurement of $\sigmav$ to derive distance ratios or deriving galaxy mass-density profiles \citep{li2018strong}.

At low and high values of each parameter, we observed a bias and we observed an expected high uncertainty. 
This likely caused by the relatively low number of training examples in these regions. 
Additionally, systems with smaller Einstein radii are likely harder to estimate due to the lack of differentiation of canonical lensing features in those systems.
More examples in these regions of parameter space could lower the uncertainty.
However, as the model confidence level improves systematic errors associated with this parameter region may also be revealed.
In future work, we seek to a) address the interplay and trade-offs for various kinds of uncertainties; and b) to  explore how changes in image quality affect these results. 


The estimations from velocity dispersion deviations were in a regime lower than $10\%$. 
This precision is competitive with spectroscopic surveys such as BELLS \citep{bolton2012boss}, and SLACS \citep{bolton2006sloan}. 
It is worth noticing that the input of our method includes not only the information from three bands, but also morphology, and strong lensed sources which are expected to be useful to constrain the velocity dispersion. 
In fact, if the $\thetae$ or angular separation $\theta$, $z_l$ and $z_s$ are well-constrained, one could estimate the velocity dispersion $\sigma_v$ with this level of accuracy, given a density profile and a cosmology \citep{davis2003strong}. 
In fact, if the velocity dispersion is constrained from strong lensing, it could be applied to modified gravity tests such as \cite{cao2017test,schwab2009galaxy}. 
This can be further investigated with techniques, such as the Local Interpretable Model-Agnostic Explanations \citep[LIME;][]{ribeiro2016model}. 
This will also be the subject of future investigation.
 
It should be noted that due to the somewhat idealized nature of $\mathrm{LensPop}$'s prescription for simulating a DES-like image dataset, the uncertainties quantified here may be lower than the uncertainty in our inferences from real imaging data. 
While $\mathrm{LensPop}$'s prescription is well-motivated by both theoretical and empirical considerations, it makes a number of simplifying assumptions about the populations of lenses and sources, as well as the simulated observing conditions of the systems (some of which are discussed in \S 7 of \citet{Collet2015}). 
Real strong lensing systems are likely to have characteristics that deviate from these assumptions to various extents. 

More crucially, $\mathrm{LensPop}$ assumes each lensing system is found in isolation. In reality, elliptical galaxies, which constitutes the majority of galaxy-scale lenses, tend to cluster, which leads to external perturbations to the lensing potential of the lens system due to nearby masses, as well as the crowding of the field-of-view near the lens system by these objects. $\mathrm{LensPop}$ also ignores objects that may be situated along the line-of-sight of the lens by coincidence. These inhomogeneous scenarios can result in higher uncertanties when the regression is applied to real data.  

In such regimes, one needs to properly address the systematic uncertainties from the data.
A possible way to reduce the impact of uncertainties in data due to factors unaccounted for in the idealized simulations might be to do transfer learning or domain adaption, in order words, start from the models presented in this paper, or parts of it, and make a fine tuning, for real data.
However, since there are orders of magnitude fewer strong lensing systems discovered in real data to date, to properly train this scheme is a major challenge and one might also need to make use of data augmentation methods. 
We are currently evaluating the relevance of idealized simulations by working on simulations with increased degrees of realism. This work therefore represents a novel step towards building a more robust framework to analyse strong lensing systems found in current and future ground-based survey data. 

It is worth mentioning a significant part of the real data might have inhomogeneous observational conditions and a the fine tuning should consider simulations with different exposure times/noise levels in different bands, or focus on real lenses with higher signal-to-noise ratios. 

\section*{Acknowledgements}

\subsection*{Author Contributions:}
\noindent\textbf{Bom}: Developed the methodology and neural netarchitectures; performed regression and uncertainty analysis; created plots; wrote and edited.\\
\textbf{Poh}: Created simulated dataset and image cut-outs; created plots; wrote and edited. \\
\textbf{Nord}: Performed analysis of results; designed diagnostics and edited document.\\
\textbf{Blanco-Valentin}: Wrote code for regression and Created plots.\\
\textbf{Dias}: Performed Linear fit and evaluated the the training.

This paper and work is supported by the Deep Skies Community\footnote{\url{https://deepskieslab.com/}}, which helped to bring together the authors and commenters.
The authors of this paper have committed themselves to performing this work in an equitable, inclusive, and just environment, and we hold ourselves accountable, believing that the best science is contingent on a good research environment.
This paper also made use of the Plot Deep Design \footnote{\url{https://github.com/cdebom/plot_deep_design}} library to make plots of the presented architecture. 

This manuscript has been authored by Fermi Research Alliance, LLC under Contract No. DE-AC02-07CH11359 with the U.S. Department of Energy, Office of Science, Office of High Energy Physics.

C.Bom would like to thank Fermilab for the financial support during his visit. 
C.Bom also would like to thank M. Makler for useful discussions. 
The authors would like to thank P. Souza Pereira for supporting this project by providing access to GPUs.

\newpage
\bibliographystyle{model2-names} 
\bibliography{bibliografia}

\end{document}